\documentclass[letterpaper,twocolumn,10pt]{article}
\usepackage{usenix-2020-09}

\usepackage{xurl}

\usepackage{pgfplots}

\usepackage{tikz}
\usetikzlibrary{arrows.meta, positioning}
\usepackage{amsmath}

\usepackage{forest}

\usepackage[inline]{enumitem}

\usepackage{lipsum}  

\usepackage{amssymb}
\usepackage{pifont}

\usepackage{stix}

\usepackage{xstring}

\DeclareRobustCommand{\agent}[1]{%
	\IfStrEqCase{#1}{%
		{A}{\textit{A}}%
		{B}{\textit{B}}%
		{C}{\textit{C}}%
		{D}{\textit{D}}%
		{E}{\textit{E}}%
		{F}{\textit{F}}%
		{G}{\textit{G}}%
		{H}{\textit{H}}%
		{I}{\textit{I}}%
		{J}{\textit{J}}%
		{CSC Global}{\textit{CSC Global}}%
		{Rijksoverheid}{\textit{Rijksoverheid}}%
		{KPN Zakelijk}{\textit{KPN Zakelijk}}%
		{SURF}{\textit{SURF}}%
		{MarkMonitor}{\textit{MarkMonitor}}%
	}[#1]%
}

\DeclareRobustCommand{\domain}[1]{%
\IfStrEqCase{#1}{%
	{verfdenachtblauw.nl}{{$\alpha$}}%
	{lachmetdedonder.nl}{{$\beta$}}%
	{breekdeschaduw.nl}{{$\gamma$}}%
	{buighetlicht.nl}{{$\delta$}}%
	{springinhetdiepe.nl}{{$\epsilon$}}%
	{breekdegolven.nl}{{$\zeta$}}%
	{verliesjeinhetniets.nl}{{$\eta$}}%
	{schitterinhetduister.nl}{{$\theta$}}%
	{buighetgeluid.nl}{{$\iota$}}%
	{dwaalindemist.nl}{{$\kappa$}}%
}[#1]%
}

\newcommand{\cmark}{\ding{51}}%
\newcommand{\xmark}{\ding{55}}%
\newcommand{\fullsquare}{\ensuremath{\blacksquare} }%
\newcommand{\partialsquare}{\ensuremath{\squareulblack} }%
\newcommand{\emptysquare}{\ensuremath{\square} }%

\begin{document}

\date{}

\title{\Large \bf Domijn: The Security of Domain Registrars and the Risk of a Domain Name Takeover}

\author{
{\rm Koen van Hove}\\
NLnet Labs \& University of Twente\\
koen@nlnetlabs.nl
\and
{\rm Jeroen van der Ham-de Vos}\\
University of Twente\\
j.vanderham@utwente.nl
\and
{\rm Roland van Rijswijk-Deij}\\
University of Twente\\
r.m.vanrijswijk@utwente.nl
} 

\maketitle


\begin{abstract}
Domain names are key assets for organisation. They anchor an organisation's online presence and reputation, and serve as linking pin for web services and, e.g., email. Consequently, a malicious takeover of a domain can lead to significant damages. Organisations register domain names through so-called registrars, a type of business that plays a key role in the domain name industry. This implies that registrars play an important part in safeguarding against malicious takeovers of domains. In this paper we empirically study how registrars implement security controls to prevent against such takeovers. We focus on the top 10 most popular registrars for the .nl ccTLD. We present the results of this study in light of a model for the impact of domain takeovers, that analyses the possible consequence of a takeover. We contrast this against the impact of two other well-known threats: ransomware and DDoS attacks. We find that all registrars in our study implement relatively effective security measures, but that they fall short in more advanced security controls, such as the proper implementation of two-factor authentication. We also find that a domain takeover can have significant impact, potentially equalling that of a ransomware attack. 
\end{abstract}

\section{Introduction}
A domain name is an important asset for 
organisations. It is where their website is, what is used to receive email, where many of their internal services live, and more. Generally an organisation registers their domain name a so-called \emph{registrar}. A registrar manages the registration at the registry 
and may also provide additional services,
such as managed DNS hosting to also provide the technical services needed to operate an Internet domain.
While registrars are the primary actor responsible for the administrative handling of domain name registrations, the actual business model is sometimes more complicated. Larger registrars often allow their services to be offered by resellers. This allows, e.g., hosting companies that themselves do not have the proper accreditation to also offer domain name registrations to their customers.

Most registrars and resellers offer their services through a web portal where information can be managed, such as: 
\begin{enumerate*}
	\item Ownership information;
	\item Data that ends up in public databases containing information about the domain name and the holder;
	\item Domain Name System (DNS) nameservers\footnote{\label{fn:dnstutorial} For an in-depth discussion of the technical aspects of the Domain Name System, we refer to a tutorial by Van der Toorn et al.~\cite{vandertoorn-tutorial-2022}.}, which tell those resolving the domain name where to find the records for that domain;
	\item Transfer codes, a code allowing someone to authorise moving a domain name to another registrar;
	\item DNSSEC keys\footref{fn:dnstutorial}, allowing someone to set the keys used to cryptographically sign the records in the DNS for this domain.
\end{enumerate*}

This setup means that a registrar's web portal is a crucial element in protecting the security of a domain name registration. If a malicious party manages to log in to the web portal on behalf of someone else, they can likely take over the domain name. 
%

Whilst we did notice several large registrars explicitly focus on brand protection and the perceived risk of a domain takeover, we noticed a lack of scientific models describing the impact of such a takeover, as well as a lack of scientific work on the technical likelihood of such a takeover. In this paper we aim to fill that gap by modelling the risk of a domain takeover for an organisation, and assessing it on using the NIST Risk Assessment Scale \cite{nist_2012} as shown in Figure~\ref{fig:nist-assessment}. We do so by first analysing the likelihood of a malicious party being able to take over a domain name by systemically investigating the technical security of registrars and resellers. Then we analyse the impact of a domain takeover by taking inspiration from existing models for two other, well-studied attacks for ransomware and DDoS. We then combine the likelihood and impact to assess the general risk.

\begin{figure*}[h]
	\begin{center}
		\centering
		\begin{tabular}{l|ccccc}
			\hline
			\textbf{Likelihood of Threat Event} & \multicolumn{5}{c}{\textbf{Likelihood Threat Events Result in Adverse Impacts}} \\
			\textbf{Initiation or Occurrence} & Very Low & Low & Moderate & High & Very High \\
			\hline
			Very High & Low & Moderate & High & Very High & Very High \\
			High      & Low & Moderate & Moderate & High & Very High \\
			Moderate  & Low & Low & Moderate & Moderate & High \\
			Low       & Very Low & Low & Low & Moderate & Moderate \\
			Very Low  & Very Low & Very Low & Low & Low & Low \\
			\hline
		\end{tabular}
	\end{center}
	\caption{The National Institute of Standards and Technology (NIST) risk assessment table} \label{fig:nist-assessment}
\end{figure*}

We focus on .nl as The Netherlands is considered a role-model according to the International Telecommunication Union's (ITU) Global Cybersecurity Index 2024 report \cite{itu_2024}, scoring 20 out of 20 in nearly every category including technical measures. The most popular registrars and resellers used for .nl are Dutch. This means our results likely paint a `best case' scenario and paint a too optimistic picture rather than a too pessimistic picture.

%

\vspace{0.8\baselineskip}
The contributions of this paper are:
\begin{enumerate}
	\item An empirical study into the security measures of domain registrars and resellers, and a characterisation of the shortcomings in the security of domain registrars and resellers, with as key finding the improper implementation of two-factor authentication;
	\item An analysis for the impact of a domain name takeover for an organisation;
	\item A risk assessment of a domain name takeover for organisations based on this empirical study and impact model.
\end{enumerate}

\textbf{Outline} -- the structure of this paper is as follows: First we describe the background and technical terminology in Section~\ref{sec:background}. After that, in Section~\ref{sec:related-work} we describe the related work. In Section~\ref{sec:methodology} we describe the methodology for which registrars and resellers we test the security of, and how we test them. In Section~\ref{sec:results} we show the results of our tests and the vulnerabilities we found. In Section~\ref{sec:impact} we analyse the impact of a domain takeover for those organisations and make the risk assessment. In Section~\ref{sec:countries} we analyse how applicable these results and impact are outside the Netherlands. In Section~\ref{sec:disclosure} we describe how we disclosed the vulnerabilities we found to the registrars and resellers. In Section~\ref{sec:discussion} we discuss what these results mean in broader context and the limitations.  In Section~\ref{sec:recommendations} we make recommendations on how to make hostile domain name takeovers less likely. Lastly, in Section~\ref{sec:conclusion} we draw our conclusions. 

Our work does not exist in a vacuum outside of society. For that reason, in Section~\ref{sec:ethics} we describe the ethical considerations and steps we have taken to minimise adverse effects.

\section{Background}\label{sec:background}
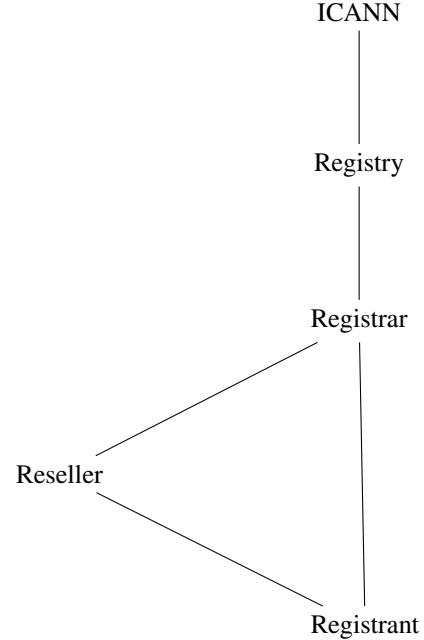
\begin{figure}[h]
	\begin{center}
		\centering
\begin{tikzpicture}[
	node distance=1.5cm and 2.5cm,
	every node/.style={draw, rectangle, align=center}
	every edge/.style={->, thick}
	]
	
	\node (icann) {ICANN};
	\node (registry) [below=of icann] {Registry};
	\node (registrar) [below=of registry] {Registrar};
	\node (reseller) [below left=of registrar] {Reseller};
	\node (registrant) [below right=of reseller] {Registrant};
	
	\draw (icann) -- (registry);
	\draw (registry) -- (registrar);
	\draw (registrar) -- (reseller);
	\draw (registrar) -- (registrant);
	\draw (reseller) -- (registrant);
	
\end{tikzpicture}
	\end{center}
	\caption{The Registry-Registrar-Reseller-Registrant relationship where ICANN is the root} \label{fig:rrrr-model}
\end{figure}

The Domain Name System (DNS) is a system to make the Internet accessible to human beings \cite{icann_2012}. Rather than having to remember a string of numbers (such as 203.0.113.5) called an Internet Protocol (IP) address to access something on the Internet, the DNS uses human readable domain names (such as example.com), and allows for quick lookup of the associated IP address, much like a phone book can be used to look up phone numbers for an organisation or name. Apart from IP addresses, the DNS is also used to relay where email for that domain name should be delivered, among other things.

The DNS is hierarchical, as shown in Figure~\ref{fig:rrrr-model}. At the top is the Internet Corporation for Assigned Names and Numbers (ICANN), they coordinate the DNS. ICANN delegate the management of the top-level domains (TLDs) to the registries -- examples of TLDs are `.nl', `.com', and `.edu'. Registries manage the list of all domain names under their TLD. Registries delegate the task of selling domain names to one or more registrars. They are the organisations where one would go to register, .e.g., `example.org'. The person or organisation who registers such a domain name is called the registrant -- they are the one who hold `example.org'. Sometimes there is another layer between the registrar and registrant called a reseller. From the perspective of the registrant, there is no difference between a registrar and reseller -- both enable you to register your domain name. The only difference is that they do not have a direct contract with the registry, but with a registrar.

In order to find the IP address for a domain name, e.g. my.example.org, your computer does the following:
\begin{enumerate}
	\item Your computer asks the root (ICANN) where to find .org, which tells you where to find .org;
	\item Your computer asks .org where to find example.org, which tells where to find example.org;
	\item Lastly your computer asks example.org where to find my.example.org, which then tells you where to find my.example.org.
\end{enumerate}

In the original version of the DNS, none of the answers were cryptographically signed. This meant that, much like the telephone game, any adversary-in-the-middle could change any of the answers. DNSSEC was added later to make answers cryptographically verifiable. The registrar plays a role in this, as it has to send the key material for a domain name to the registry \cite{chung_2017}, so that when one asks the .org registry where to find `example.org', the registry can answer both where to find example.org, but also which key answers from example.org (e.g. for my.example.org) will be signed with.

%

\subsection{Terminology}
We believe it is worthwhile to provide a brief description of the technical terms used throughout this paper:
\begin{description}
	\item[TLD] A Top-Level Domain (TLD) is a domain in the DNS at the highest hierarchical level. Examples are .com or .nl.
	\item[Domain name] The part of a network address that defines a realm of control under a TLD, such as example.com or voorbeeld.nl.
	\item[DNS] The Domain Name System (DNS) as defined in RFC 1035 \cite{rfc1035} and subsequent standards are the set of protocols defining the technical workings of domain names.
	\item[Resolver] A resolver retrieves information (records) associated with the domain name, such as the IP address it can find a server, or which server email should be sent to.
	\item[Nameserver] A nameserver contains the information about the records for a domain name, such as the IP address it can find a server, or which server email should be sent to, and provides that information when a resolver asks for it.
	\item[DNSSEC] DNS Security Extensions (DNSSEC) \cite{rfc9364} are a set of protocols that provide origin authentication of DNS data, allowing the resolver to check DNS data has not been tampered with.
	\item[Registry] A registry is the operator of a TLD, who manages all the database of all domain registrations within that TLD. \cite{icann_2017}
	\item[Registrar] A registrar is an organisation accredited by a registry to sell domain registration services to the (general) public. They may, but need not, provide more services.
	\item[Reseller] A reseller is an organisation that sells domain registration services to the (general) public via a registrar.
	\item[Registrant] A registrant is an organisation or individual that buys and holds the domain name.
	\item[WHOIS] WHOIS (not an acronym, pronounced as `Who is') is a protocol with which information about the registrant can be retrieved about an internet resource such as a domain name \cite{rfc3912}. Information can include administrative and technical names, addresses, phone numbers and email addresses.
	\item[RDAP] The Remote Data Access Protocol (RDAP) is the follow-up protocol from WHOIS \cite{rfc7483}. It is used for the same purpose, but retrieves it in a standardised format.
	\item[DDoS attack] A Distributed Denial of Service (DDoS) attack is a type of cyber attack where many systems try to flood a service, making it unavailable. \cite{razavi_2023}
	\item[Ransomware] Ransomware is malware that encrypts a users' files, effectively holding data and systems hostage until an amount of money (the ransom) is paid to the parties.
\end{description}

\section{Related Work}\label{sec:related-work}
Due to the nature of this paper, we split our related work up in three parts: 
\begin{enumerate}
\item The first part is about the related academic work regarding the technical aspect of domain names and possible ways to abuse aspects of its operation;
\item The second part is about the related academic work on impact models for cyber attacks;
\item The third part is about instances of domain name takeovers we found in news articles.
\end{enumerate}

\subsection{DNS and Registrar Attacks}
DNS and DNS vulnerabilities have been studied in great detail, for a good overview we refer to the survey by Schmid~\cite{schmid_2021}.

DNSSEC, its working, security, and deployment, have been studied greatly. We refer to the work by Chung et al. \cite{chung_2017} and Lian et al. \cite{lian_2013} for an overview. Registrars hold an important position in DNSSEC, as they are the ones who can add, change, and remove the DNSSEC keys used for a domain name. Hence we focus on the role of the registrar. The UK National Cyber Security Centre published guidance for good security practice for domain registrars \cite{ncsc_2025}.

When it comes to the registrar's role, Chung et al.~\cite{chung_2017} studied the role registrar's play in DNSSEC. They find a worrying lack in security posture, e.g., demonstrating that some providers accept changes to key material via email without checking if these mails come from the legitimate domain name owner. Akiwate et al. \cite{akiwate_2021} explored a renaming scheme some registrars use to handle expired domain names where those domain names are still referred to by another domain name, making those other domains susceptible to abuse. 

Vissers et al. \cite{vissers_2017} do something similar by looking at what they refer to as `nameserver typosquatting'. The DNS allows for domain names to refer to other domain names, similar to how a dictionary can refer to another word for the definition of a word. Unlike a dictionary however, the DNS allows referring to more than one other domain name to allow for redundancy. Vissers et al. analyse where typos have been made in one of these names, such that, if one were to register that domain, they could reply with malicious answers.

Zhang et al. \cite{zhang_2024} do something similar with stale `glue' records. Glue records are necessary when the hierarchy is self-referential. For example, .org might answer that the data for example.org can be found at ns1.example.org. We cannot ask example.org for the data for ns1.example.org as we do not know where example.org can be found yet. In these cases, a `glue' record allows .org to also inform us where we can find ns1.example.org directly.

Schlamp et al. \cite{schlamp_2015} look at domain names that have expired that are listed in internet resource databases, which shows the holder information for those resources. By registering those domain names, one can claim ownership and take over those resources. 

We could find no research that systematically looks into the security measures taken by domain registrars, although Akiwate et al. \cite{akiwate_2021} do show that certain registrars are disproportionately associated with malicious domains, and Munny et al. \cite{munny_2025} show that certain providers are used more commonly for toll scams.

\subsection{Impact Models}
In order to understand the risk of a domain name takeover, we need both the likelihood and the impact of such an event.

There are impact models and quantification for other kinds of cyber attacks. Khan et al. \cite{khan_2010} design a generalised model for quantifying cyber security. The National Institute of Standards and Technology (NIST) have also created a framework for conducting cyber risk assessments \cite{nist_2012}. Wolthuis et al. \cite{wolthuis_2021} show a model in action by using a DDoS attack as an example. 

Risk has also been quantified from a financial perspective. Razavi et al. \cite{razavi_2023} analyse the financial loss for DDoS attacks at banks, Woods et al. \cite{woods_2021} look at the impact of a company's stock value, and Gomez et al. \cite{gomez_2023} who look at how much money ransomware groups earned by analysing Bitcoin transactions.

We could not find a threat model if a registrar or reseller is breached and a domain name is taken over, although it is named by Schmid \cite{schmid_2021} and in a report by ICANN \cite{icann_2015}.

\subsection{In the News}
There are known instances of domain name takeovers. Davis \cite{davis_2025} describes a breach at a registrar using social engineering, which allowed the attacker access to the organisation's entire cloud infrastructure. Cointelegraph \cite{cointelegraph_2025} describes another breach where domain hijacking was used to steal cryptocurrency. The Register describes a breach that impacted Twitter.co.uk and the New York Times \cite{chirgwin_2013}. The Cybersecurity and Infrastructure Security Agency (CISA) mention that there has been an increase in malicious activity in DNS infrastructure \cite{cisa_2019}, and does mention verifying DNS registrar accounts, which seems to be related to a report from Cisco \cite{cisco_2018} from around the same time, where email traffic for the Lebanese and United Arabic Emirates (UAE) was redirected.
\vspace{0.8\baselineskip}

In this paper we thus aim to fill the gap by investigating the state of registrar and reseller security, and analyse the impact by creating an impact model inspired by already quantified and modelled threats. This way, we can assess the risk of a domain name takeover.

\section{Methodology}\label{sec:methodology}
As described in Section~\ref{sec:background}, for the registrant there is little difference between a registrar and reseller. In most cases the registrant will log into the web portal provided by where they bought their domain name, unaware of whether they are a registrar or reseller. A change made by a registrant at the reseller will still end up at the registry in the same way as it would when making that change at a registrar. For that reason, \textbf{we group these registrars and resellers and refer to them as `agents'}. 

\subsection{Determining Agents}\label{sec:determining-agents}
We use the top 1,000,000 domains from Cloudflare Radar between 22-29 July 2024, which they base on data from their public DNS resolver \emph{1.1.1.1}. We filter out the .nl domains -- this results in just over 9000 .nl domains. We elaborate why we focus on .nl in Section~\ref{sec:countries}. Based on this we aggregated the most common agents by querying for their registrar and reseller data over RDAP. This can be seen in figure \ref{fig:top-registrars}. 

Some agents allow for sign-ups without human interaction, mainly through a web portal. \agent{A} has two distinct registrar/reseller names they use in the WHOIS data, but both use the same portal. \agent{KPN Zakelijk} has a sign-up process with specific requirements we did not meet. 

Not all registrars that only act as portals for resellers (i.e. do not do registrations from customers directly) always accurately include the reseller information in the WHOIS data (such as Registrar.eu, Key-Systems GmbH, and The Registrar Company B.V.), and these have been excluded. 

The agent landscape in The Netherlands (where most .nl domains are used) is varied. Over 300 agents are required to reach over 90\% market share. In total we noticed over 1000 different agents within these 9000 domain names. We believe this selection provides an accurate view of agents that are commonly being used for .nl domain names. Based on the names from the RDAP data, most agents appear to be Dutch. 

We look at the top 10 agents from this set that allow for `automated' sign-up, meaning by entering our details on a website and spending a small fee. The others, such as \agent{CSC Global}, require an onboarding process through a sales representative, which would require us to have a business at scale, or are closed for signing up entirely, such as \agent{SURF}, which only provides services to its members.

Even though the national government (`Rijksoverheid') is its own registrar, some organisations within the government use a different registrar. When checking the WHOIS data of the list of national government domains, all agents listed in Figure~\ref{fig:registrations} appeared at least once (though often several times).

\begin{figure}[h]
	\begin{center}
		\resizebox{\columnwidth}{!}{
		\begin{tabular}{|l|c|c|c|}
			\textbf{Agent} & \textbf{Market share} & \textbf{Automated} & \textbf{Country}  \\ 
			\hline
			\agent{A}*         & 20.31\%               & \cmark         & NL           \\
			\agent{B}          & 3.05\%                & \cmark         & NL           \\
			\agent{C}            & 2.47\%                & \cmark         & FR           \\
			\agent{D}          & 2.37\%                & \cmark         & NL           \\
			\agent{A}*         & 1.98\%                & \cmark         & NL           \\
			\agent{E}           & 1.93\%                & \cmark         & NL           \\
			\agent{F}      & 1.92\%                & \cmark         & NL           \\
			\agent{CSC Global}       & 1.70\%                & \xmark         & US           \\
			\agent{Rijksoverheid}    & 1.53\%                & \xmark         & NL           \\
			\agent{KPN Zakelijk}     & 1.23\%                & ~~~~\cmark**   & NL           \\
			\agent{SURF}             & 1.14\%                & \xmark         & NL           \\
			\agent{G}           & 1.14\%                & \cmark         & NL           \\
			\agent{H}           & 1.07\%                & \cmark         & DE           \\
			\agent{I}              & 1.07\%                & \cmark         & NL           \\
			\agent{MarkMonitor}      & 0.91\%                & \xmark         & US           \\
			\agent{J}       & 0.87\%                & \cmark         & NL          
		\end{tabular}
		}
	\end{center}
	\caption{The top .nl registrars and resellers based on the Cloudflare Radar data from 22-29 July 2024 based on data from their public DNS resolver 1.1.1.1. } \label{fig:top-registrars}
	
\end{figure}

\subsection{Registering Domains}
We determined which agents to look at. As we also highlight in Section~\ref{sec:ethics}, we only use our own domains for testing. For that reason, we have registered the domains as a natural person at each agent. We did so using a real, albeit not colloquially used, name, and an address on the 
University of Twente. 
We used a newly created AOL email address, and a new phone number specifically for this purpose.

\begin{figure}[h]
	\includegraphics[width=1\linewidth]{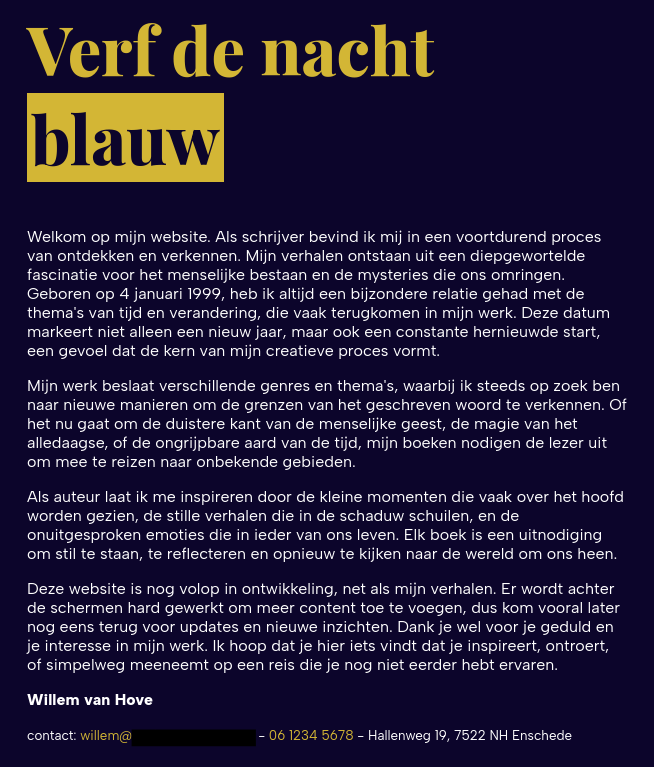}
	\centering
	\caption{A screenshot of the website we created for the fictional writer. The domain name has been redacted}
	\label{fig:verfdenachtblauw}
\end{figure}

\subsection{Authentication by Agent}
For each agent we registered a domain with, we kept track of the information they asked of us during registration, as this can later be used to authenticate us. This can be seen in Figure~\ref{fig:registrations}. In all but one case this was only the name, email address, telephone number, and address. We also kept track of which two-factor authentication options were provided, and whether it was mandatory to have one. Timed One-Time Passwords (TOTP) seem to be far the most popular method of multi-factor authentication.

\subsection{Information Available}\label{sec:information-available}
We created a website (as seen in Figure~\ref{fig:verfdenachtblauw}, name changes per domain) for a fictional book writer that lists that it is still under construction. The website contains some text, and the author's email address specific to that domain, telephone number, and physical address. 

Out of the 10 agents we registered a domain with, 8 show our email address we registered with as administrative and/or technical contact in the public WHOIS data. Only \agent{C} and \agent{H} do not show the address, but instead use their own address. However, this is likely to change in the future due to new policy from Stichting Internet Domeinregistratie Nederland (SIDN) since October 2023 \cite{sidn_2023}. Most other data is redacted for privacy reasons. \agent{D}, \agent{J}, \agent{I}, \agent{G}, and \agent{F} use the email address that was used for sign-up as login name and email address in the WHOIS data by default.

The data the registrar has about the registrant is listed in Figure~\ref{fig:registrations}. We assume one is able to find out the following information available about the holder of the domain (e.g. through open source intelligence):
\begin{enumerate}
	\item The domain name;
	\item The name of the holder;
	\item The email address associated with the account;
\end{enumerate}

\subsection{Experiment Setup}
We have determined which agents to look at, what data they possess about the registrant, and what data an outsider might reasonably have about the registrant. Based on this, we systematically test, for each of the agents:
\begin{enumerate}
	\item How likely it is that we are able to find a working password for a specific user;
	\item What technical measures they use to protect against password / TOTP brute forcing, by testing 100 TOTP codes in quick succession;
	\item What verification questions are asked when trying to get the email address associated with an account changed by claiming the password reset email does not arrive via a phone call;
	\item Whether based on the information an outsider has available, they are able to make unauthorized modifications to a domain name they do not hold.
\end{enumerate}

\subsection{Ethical Considerations}\label{sec:ethics}
From the start of this research, we involved the ethics commission of the 
University of Twente. 
We always made sure to only do testing using our own resources and own domains. The registrars and resellers where issues were found were informed before publication as described in Section~\ref{sec:disclosure}. In all our testing we made sure to minimise the impact on existing customers of the registrars and resellers. The registry, Stichting Internet Domeinregistratie Nederland (SIDN) was informed beforehand.

We made very sure to not put the blame or responsibility on a specific person or to guilt-trip or pressure them to do something for us. Their names and identifying information were not recorded.

We made the conscious decision to not publish the names of the registrars and resellers that we interacted with. We are aware that their names can be deduced from the data, and of the risk it poses to these organisations and their customers, but we believe it is infeasible to do this research in a way that cannot be traced back to the registrars and resellers. We believe that there is a greater public benefit by publishing this paper than by withholding it. 

\section{Results}\label{sec:results}
\subsection{Password Availability}\label{sec:password-availability}
An estimated 43\% to 51\% of users reuse their passwords \cite{das_2014}. According to Ablon et al., 43\% received a breach notification in their lifetime \cite{ablon_2016}. Research at Google discovered that out of password stolen or leaked online, between 7\% and 25\% matched the Google account's login data \cite{thomas_2017}. Using these numbers we can obtain a rough estimation for the number of passwords being available.

Out of the 9000 .nl domains, we identify just over 7500 unique email addresses in the WHOIS data. We checked for these email addresses whether they had any data in Have I Been Pwned \cite{haveibeenpwned}. We found breach notifications for roughly 54\% of those email addresses. Of those accounts with breach notifications, 56\% had a breach that involved the leaking of passwords. These numbers also include `privacy' email addresses as provided by, for example \agent{C}, which are unique and unlikely to be used anywhere else. It is noticeable that a lot of these domains use an administrative and/or technical contact that uses a private email address. For example, around 5\% uses an @gmail.com address.

Combining these statistics, we believe that between 2\% and 8\% of domains currently have a working password available online from a leak. We have not tried to confirm this number.

\subsection{TOTP Bypass}
We can approach the time it takes to brute force the TOTP token using the binomial distribution. We assume the current TOTP code, the previous code, and the next code are valid at any point in time. We fire off around 100 requests per second. By running this for an hour, our probability to correctly guess the TOTP code becomes: $$1 - B(100 \cdot 3600, \dfrac{3}{1,000,000}) \approx 55\% $$

We therefore check whether we can burst 100 requests in one second -- if the agent does not prevent our attempts, we could likely brute force the TOTP. We do this by logging in as usual, sending off 100 TOTP requests, and then trying to log in using the correct TOTP code on the original page. If the correct TOTP code still works, it gives a strong indication no rate-limiting is being applied. If the correct TOTP code does not work, we try changing our IP address and entering the code again. If that does work, it gives us a strong indication that rate limiting is only applied per IP address, and not per account. We only burst 100 requests to limit the possible damage on the agent's servers -- also see Section~\ref{sec:ethics}. Our findings as tested in the end of 2024 and early 2025 are shown in Figure~\ref{fig:totp-protection}. 

RFC 4226 section 7.3 recommends several throttling techniques for one-time passwords \cite{rfc4226}, but it states they have to be applied across login sessions. Only \agent{A} and \agent{I} did this on a per-account basis. \agent{D} did lock out our IP address for half an hour but not our account. \agent{E} returned a ``429 Too Many Requests'' page. In both cases changing our IP address and reloading the page allowed us to sign in. \agent{J} did not tell us that we were rate limited, but refused our correct TOTP token with a 429 error. However, after changing our IP address we were also let in.

To our knowledge organisations such as the National Institute of Standards and Technology (NIST) and the National Cyber Security Centre (NCSC) provide no guidelines for this. The Open Worldwide Application Security Project (OWASP) does not mention rate limiting, but does mention alerting the user at a failed two-factor authentication login attempt \cite{owasp_2026}. None of the agents we tested did this.

\begin{figure}[t]
	\begin{center}
		\begin{tabular}{|l|c|l|l|}
			\textbf{Agent} & \textbf{BFP} & \textbf{BFP method} & \textbf{Reset} \\ 
			\hline
			\agent{A} & \fullsquare & Lifetime expires & Text message \\
			\agent{B} & \emptysquare &  & Contact support \\
			\agent{C} & \emptysquare &  & Recovery codes \\
			\agent{D} & \partialsquare & IP rate limiting & Recovery code \\
			\agent{E} & \partialsquare & IP rate limiting & Contact support \\
			\agent{F} & \emptysquare &  & Contact support \\
			\agent{G} & \emptysquare &  & Contact support \\
			\agent{H} & \emptysquare &  & Contact support \\
			\agent{I} & \fullsquare & Session expires & Contact support \\
			\agent{J} & \partialsquare & IP rate limiting & Contact support 
		\end{tabular}
	\end{center}
	\caption{The agents, their TOTP brute force protections (BFP) -- if applicable -- and reset mechanism in case the TOTP token was lost. \fullsquare means the agent implemented rate limiting on a per-account basis, \partialsquare means rate limiting was implemented, but not per account but per IP address, and \emptysquare means no rate limiting was implemented.} \label{fig:totp-protection}
\end{figure}

\subsection{Helpdesk Calls}\label{sec:phone-calls}
Phone calls involve another person directly. Hence for that reason we have taken the utmost care to not harm any of the operators in the process. See Section~\ref{sec:ethics} for more details.

We created a phone script to mimic someone calling to reset their password, and not receiving the reset email. The information available to the person calling is:
\begin{enumerate}
	\item the domain name of the registrant;
	\item the name of the registrant;
	\item the email address of the registrant;
\end{enumerate}
This is in line with the data we considered to be available described in Section~\ref{sec:information-available}. The caller then tries to either receive the (new) password via phone or get the email send to their own email address, thereby gaining access to the account. We created a rough call flow diagram as shown in Appendix~\ref{sec:flowchart} -- we consider it infeasible to account for every possible answer.

Our goal is not to come up with the most persuasive social engineering scenario, but rather determine what information is required to gain access to an account. Our hypothesis is that with fairly limited data about a domain name, access can be acquired, similar to what Chung et al. \cite{chung_2017} found. The information available to the operator on the other end about us is limited. The information we provided, as can also be seen in Figure~\ref{fig:registrations}, tends to be not particularly secret. Additional information they have consists of things such as receipts and invoices, or asking for ID. 

We do want to explicitly point out that simple measures exist to prevent this kind of impersonation. Calling the phone number on file, sending an email to the known address, or sending new login details per letter to the registered address, are all first-line defence options to prevent account hijacking. In all of these cases, the legitimate account holder would notice the attempt.

\agent{A} does not have a phone number. Their system is outlined in their knowledge base, and involves a copy of an identity document and copy of payment, as well as a business register extract. 
\agent{C}, \agent{J}, and \agent{D} also state they ask for proof of ID via email. We have thus excluded these from this test. We hid the phone number we called from.

\agent{B} and \agent{F} share a customer service team -- we only called \agent{B}. \agent{E} does not do phone support nor does their website list how to do it, the only option is to open a support ticket.

We list the information that was asked of us in Figure~\ref{fig:data-asked}. None of our attempts resulted in gaining access to the account. We were not notified of any attempt by any of the registrars either.

\begin{itemize}
	\item \textbf{\agent{B}} asked for the name, domain, customer number (which we claimed to not have), and then asked us to send a copy of our ID card in a response to their email, as well as the new email address;
	
	\item \textbf{\agent{G}} asked us to send a redacted copy of our ID card (with only the name visible) to their general support email address;
	
	\item \textbf{\agent{H}} could not do anything without a customer number;
	
	\item \textbf{\agent{I}} asked us to call back from the phone number known in their systems.
\end{itemize}

These results are in contrast with the results from Chung et al. in 2017 \cite{chung_2017} where a lot could be arranged via a phone call. 

\begin{figure}[h]
	\begin{center}
		\resizebox{\linewidth}{!}{
			\begin{tabular}{|l|c|c|c|c|c|c|}
				\textbf{Agent} & \textbf{Phone} & \multicolumn{5}{c|}{\textbf{Information requested}} \\
				& & \emph{Name} & \emph{Domain} & \emph{Cust. No.} & \emph{Ident.} & \emph{Tel. no.} \\ 
				\hline
				\agent{A} & \xmark & \multicolumn{5}{c|}{Website asks to send identification} \\
				\agent{B} & \cmark & \cmark & \cmark & \cmark & \cmark & \xmark \\
				\agent{C} & \xmark & \multicolumn{5}{c|}{Website asks to send identification} \\
				\agent{D} & \cmark & \multicolumn{5}{c|}{Website asks to send identification} \\
				\agent{E} & \xmark & \multicolumn{5}{c|}{Unknown}\\
				\agent{F} & \cmark & \multicolumn{5}{c|}{Same as \agent{B}}\\
				\agent{G} & \cmark & \cmark & \cmark & \cmark & \cmark & \xmark \\
				\agent{H} & \cmark & \cmark & \xmark & \cmark & \xmark & \xmark \\
				\agent{I} & \cmark & \cmark & \cmark & \cmark & \xmark & \cmark\\
				\agent{J} & \xmark & \multicolumn{5}{c|}{Website asks to send identification}
			\end{tabular}
		}
	\end{center}
	\caption{The information requested over the phone when we called the customer service of our ten agents.}\label{fig:data-asked}
\end{figure}

\begin{figure*}[h]
\begin{center}
		\begin{tabular}{|l|c|c|c|c|c|c|c|c|c|c|c|}
			\textbf{Agent} & \multicolumn{6}{c|}{\textbf{Registration details}} & \multicolumn{5}{c|}{\textbf{Two-factor authentication}} \\
			& \emph{Name} & \emph{Email} & \emph{Tel. no.} & \emph{Address} & \emph{D.o.B.} & \emph{Sec. Q.} & \emph{Email} & \emph{Phone} & \emph{TOTP} & \emph{FIDO} & \emph{mandatory}\\ 
			\hline
			\agent{A} & \cmark & \cmark & \cmark & \cmark & \xmark & \xmark  & \xmark & \cmark & \cmark & \xmark & \xmark \\
			\agent{B} & \cmark & \cmark & \cmark & \cmark & \cmark & \xmark  & \xmark & \xmark & \cmark & \xmark & \xmark \\
			\agent{C} & \cmark & \cmark & \cmark & \cmark & \xmark & \xmark & \cmark & \xmark & \cmark & \cmark & \cmark \\
			\agent{D} & \cmark & \cmark & \cmark & \cmark & \xmark & \xmark & \xmark & \xmark & \cmark & \xmark & \xmark \\
			\agent{E} & \cmark & \cmark & \cmark & \cmark & \xmark & \xmark & \xmark & \xmark & \cmark & \xmark & \xmark \\
			\agent{F} & \cmark & \cmark & \cmark & \cmark & \xmark & \xmark & \xmark & \xmark & \cmark & \xmark & \xmark \\
			\agent{G} & \cmark & \cmark & \cmark & \cmark & \xmark & \xmark & \xmark & \xmark & \cmark & \xmark & \xmark \\
			\agent{H} & \cmark & \cmark & \cmark & \cmark & \xmark & \xmark & \xmark & \xmark & \cmark & \xmark & \xmark \\
			\agent{I} & \cmark & \cmark & \cmark & \cmark & \xmark & \xmark & \cmark & \xmark & \cmark & \xmark & \cmark \\
			\agent{J} & \cmark & \cmark & \cmark & \cmark & \xmark & \xmark & \xmark & \xmark & \cmark & \xmark & \xmark 
		\end{tabular}
\end{center}
\caption{We registered domain names at each organisation from Figure \ref{fig:top-registrars}, and kept track of the information they require during the registration. From left to right: name, email address, telephone number, address, date of birth, and security questions, as well as its two-factor authentication options. We exclude protections only applied to high-risk IP addresses (such as public Virtual Private Networks (VPNs)), like typing in a code sent to an email or selecting all fire hydrants, as these can be circumvented by, e.g., using a residential IP address} \label{fig:registrations}
\end{figure*}

\begin{figure*}[h]
	\includegraphics[width=1\linewidth]{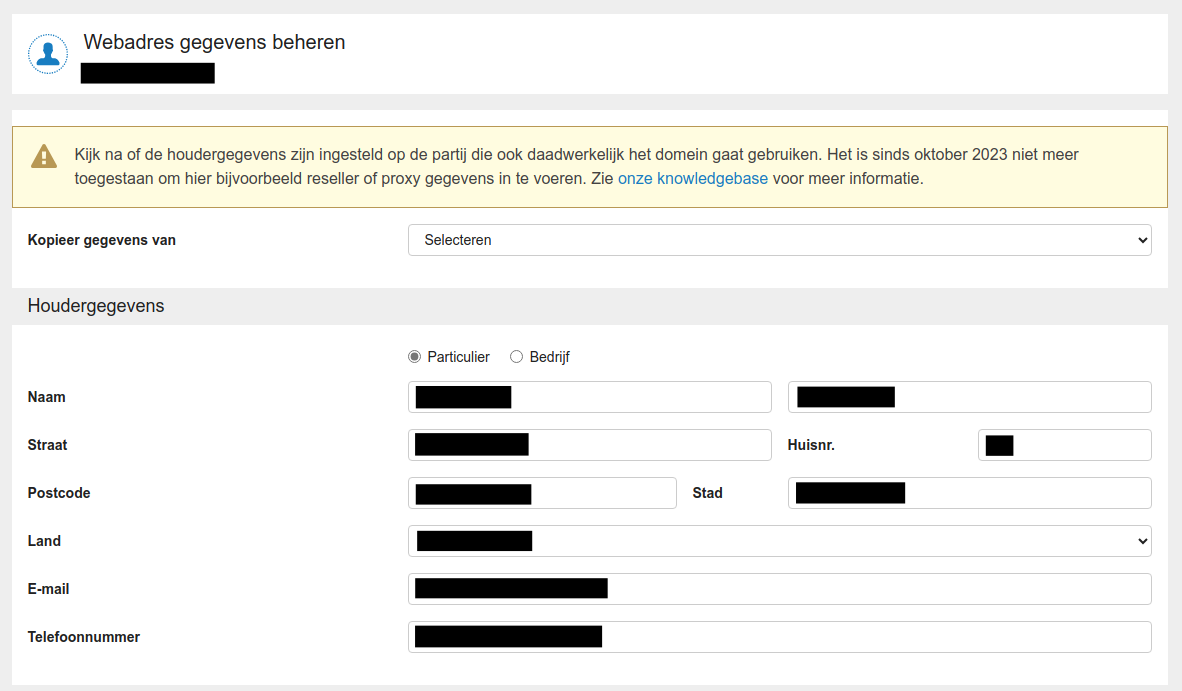}
	\centering
	\caption{The information \agent{A} shows as account holder data. From top to bottom: 1) Name, 2) Street / House no., 3) Postal code / City, 4) Country, 5) E-mail address, and 6) Phone number.}
	\label{fig:transip}
\end{figure*}

\subsection{Analysis of the Most Popular Domains}\label{sec:most-popular-domains}
In Section~\ref{sec:methodology} we describe that we look at the top 1,000,000 domains from Cloudflare Radar. We also specifically look at the top 10\% of those domain names (i.e. the top 100,000) -- of which there are 480 .nl domain names.

As can be seen in Figure~\ref{fig:top-registrars}, not all registrars make it possible to register a domain name through a web interface without human interaction. The most prominent ones are \agent{CSC Global}, \agent{Rijksoverheid}, \agent{KPN Zakelijk}, \agent{SURF}, and \agent{MarkMonitor}. These five can again be split into two:

\textbf{Private registrars} -- \agent{Rijksoverheid} is the registrar for the Dutch national government and \agent{SURF} is the registrar for Dutch education and research institutions. Access is only for their `members' (i.e. institutions and departments within their scope) -- they are thus fully closed off;

\textbf{Corporate registrars} -- \agent{CSC Global}, \agent{KPN Zakelijk} and \agent{MarkMonitor} are companies whose main service is ``protecting your brand''. They promote their services by claiming that they protect your digital assets (of which domain names are part), and that there is personalised around the clock support. They allow anyone to sign up (given they bring enough money), but only through human interaction (e.g. a sales representative).

When we analyse the 480 .nl domains names in the top 100,000 domain names from Cloudflare Radar between 22-29 July 2024 we see a shift towards these private and corporate registrars compared to Figure~\ref{fig:top-registrars}. 

We considered every domain of these 480 by hand, and find that many of the domain names used as `brand' by a major institution (i.e. the main domain an organisation is known for) are registered at either a private or corporate registrar. Out of those that are not, most are registered at \agent{A}. The domains from main institutions that are registered at neither \agent{A} nor a private/corporate registrar are the exception. 

\section{Impact of a Domain Takeover}\label{sec:impact}
In the previous sections we looked at the likelihood of being able to gain access to an agent's web portal. The question remains what one can actually do once access has been gained into the web portal of an agent, and what the impact of that is.

Stichting Internet Domeinregistratie Nederland (SIDN) introduced .nl Control \cite{sidn_2025}. With .nl Control, SIDN will call to verify the assigned person at SIDN to verify the changes, and later also require a signature before any changes are processed. This greatly limits the changes anyone can make. However, we are not aware of how wide-spread this is, but based on the list of supported providers listed on SIDN's website, none of the providers from Figure~\ref{fig:registrations} support it yet. We thus believe that most .nl domains do not use this yet. 

Without .nl Control in place, the registrant will receive an email on updates regarding the domain name, but the update will still go through nearly instantly. Once logged in, an attacker can generally do four main things:

\begin{enumerate}
	\item The attacker can request a transfer code. This enables an attacker to transfer the domain, thereby fully taking control of a domain name (and taking control away from the original holder);
	\item The attacker can change the DNS name servers to their own. This enables an attacker to (selectively) change responses to DNS requests;
	\item The attacker can view and change the holder information, such as phone numbers and addresses;
	\item The attacker can disable or change the DNSSEC keys.
\end{enumerate}

With control over the DNS responses, an attacker can do many things. This list is not exhaustive:

\begin{enumerate}
	\item Control what users see on the website;
	\item Capture the cookies of users, thereby being able to impersonate that user on the website;
	\item Obtain data victims enter on the attacker's website served on the original domain;
	\item Receive all the email for that domain, including password reset emails;
	\item Log in to any external systems that can be accessed after requesting a password reset, such as payroll and other administrative systems, as well as external cloud infrastructure;
	\item Request Transport Layer Security (TLS) certificates for the domain, making the connecting appear secure in e.g. browsers;
	\item Provide false responses to Application Programming Interface (API) calls, manipulating the data retrieved by e.g. third party integrations;
	\item In the case of Internet of Things devices, publish and push updates to end-user devices;
\end{enumerate}

When an attacker can control the DNS responses, they effectively have `the keys to the castle'. This is comparable to other digital threats an organisation might face. We pick two we consider close when it comes to impact:

\subsection{Comparison with Ransomware}
Another threat for an organisation is ransomware. Ransomware causes losses of billions of dollars \cite{oz_2022}. Ransomware is a subset of malware that blocks the user or organisation out of their own systems and/or files. In many cases it encrypts the files on the system, requiring a ransom to be paid to retrieve the key. This amount can be millions of dollars \cite{zimba_2019}. 

We argue that there are a lot of parallels to be drawn between the ransomware threat and the threat of a domain takeover, although there are differences as well:

\begin{enumerate}
	\item Both grind operations in a lot of organisations to a halt. Systems will become unavailable (e.g. due to Single Sign-On (SSO), email, etc. being unavailable);
	\item Both have the attacker as party who can solve the issue they created, though in the case of a domain takeover the registry has authority as well;
	\item Ransomware is destructive by encrypting files, domain takeovers only get rid of the reference (e.g. a local DNS resolver that overrides the takeover would still allow internal services to keep operating);
	\item Both allow for access to e.g. email (in the case of a domain takeover only to email that is received after the takeover), allowing for things like password resets in (remote) systems and access to potentially sensitive data;
	\item Both allow an attacker to perform destructive actions on third-party platforms, such as Software-as-a-Service (SaaS);
	\item Both allow an attacker to exfiltrate data from third-party platforms, such as Software-as-a-Service (SaaS), and extort the victim to prevent publication of secret or sensitive data;
\end{enumerate}

\subsection{Comparison with DDoS Attacks}
DDoS attacks can make an organisation's digital infrastructure (e.g. website, email) unavailable for the duration of the DDoS attack. Like ransomware, we believe parallels can be drawn between the threat of a DDoS attack and the threat of a domain takeover.

\begin{enumerate}
	\item Both have the potential to grind operations within an organisation to a halt. Things like email, Single Sign-On (SSO), become unavailable for the duration of the DDoS attack, or until the domain takeover is sorted;
	\item In both cases the attacker is the party who can solve the issue, either by giving back control of the domain name, or stopping the DDoS attack;
	\item DDoS attacks require the attacker to keep attacking the organisation, once they stop the services become available again. A domain takeover does not require this;
	\item A DDoS attack does not allow the attacker access to sensitive data, whereas a domain takeover can result in receiving email for the organisation, including possibly password reset emails, allowing them access to sensitive systems.
\end{enumerate}

\subsection{Relative Impact}
When we look at ransomware, we see that the impact of a domain takeover is likely lower. When we look at a DDoS attack, we see that the impact is likely higher. We thus believe that these two examples provide appropriate upper and lower bounds. This allows us to estimate the impact of a domain takeover.

We believe that this can be modelled in similar ways to the DDoS attacks described by Wolthuis et al. \cite{wolthuis_2021} and ransomware attacks from Gomez et al. \cite{gomez_2023}. We believe that there is reason to take the risk of a domain takeover into account in a similar vein to DDoS and ransomware attacks.

Combined with our findings from Section~\ref{sec:results}, we assess the likelihood of a threat occurring to be on the lower side of the spectrum, but the likelihood of adverse impacts on the higher end of the spectrum, based on the NIST assessment from Figure~\ref{fig:nist-assessment}. 

The precise impact depends too much on the organisation to make a clear assessment. For example, for an organisation providing publicly available weather data, the risk of data manipulation or data not being available might be far greater than data being leaked. Similarly, for a psychiatrist, confidentiality might be far more important than the availability of their systems. Due to the varying nature of the various kinds of attacks for various organisations, we believe we cannot make general statements about where in the risk table DDoS attacks, ransomware, and domain takeovers would land.

\section{Comparison with Other Countries}\label{sec:countries}
We aim to analyse the best-case scenario. .nl has a diverse market of agents, and a lot of them are based in the Netherlands and used for things related to the Netherlands. This allows us to reasonably use .nl to approximate the state in the Netherlands. The moderate risk we find as described in the previous section is based on analysis from .nl, and the Netherlands is a role model country according to the ITU cybersecurity index \cite{itu_2024}. The Netherlands scores 20 out of 20 in legal measures, technical measures, organization measures, and cooperation measures. Only capacity development is 19.22 out of 20. We hence believe that the risk assessment is skewed towards painting a `too positive' picture rather than a `too negative' picture.

We analyse the state of agents, not a few agents specifically. The Netherlands has a competitive registrar market, and this applies to most countries in Europe. CENTR, the association of European country code top-level domain (ccTLD) registries, writes in a report that ``The distribution networks of registrars selling European ccTLDs are on average considered competitive based on a widely used measure of market concentration'', based on the calculation of the Herfindahl-Hirschman Index (HHI) over 11 CENTR member registries in 2023 \cite{centr_2024}. According to CENTR, in 2024 the median Herfindahl-Hirschman Index (HHI) for registrars under CENTR member registries was 1324. This suggests that registrar channels for most member registries are considered competitive (i.e. have low concentration). 17 members have competitive (HHI under 1500) registrar markets and a further 9 are considered moderately concentrated. There are no member registry's with highly concentrated registrar channels.

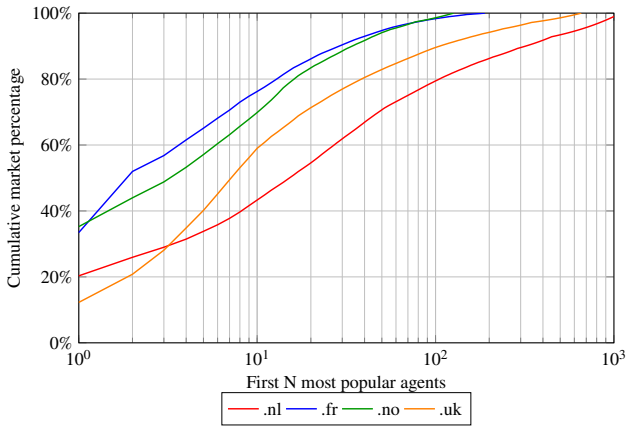
\begin{figure}[h]
\begin{center}
	\resizebox{\linewidth}{!}{
	\begin{tikzpicture}
		\begin{axis}[
			xmode=log,
			yticklabel={\pgfmathprintnumber\tick\%},
			xmin=1, xmax=1000,
			ymin=0, ymax=100,
			xlabel={First N most popular agents},
			ylabel={Cumulative market percentage},
			legend style={at={(0.5,-0.15)}, anchor=north, legend columns=4},
			width=12cm,
			height=8cm,
			grid=both,
   			legend cell align=left,
			]
			
			\addplot[red, thick] 
coordinates { (1, 20.32) (2, 25.94) (3, 28.99) (4, 31.46) (5, 33.82) (6, 35.8) (7, 37.75) (8, 39.69) (9, 41.61) (10, 43.31) (11, 44.84) (12, 46.3) (13, 47.53) (14, 48.67) (15, 49.82) (16, 50.88) (17, 51.95) (18, 52.86) (19, 53.73) (20, 54.58) (21, 55.44) (22, 56.28) (23, 57.13) (24, 57.9) (25, 58.64) (26, 59.32) (27, 60.0) (28, 60.67) (29, 61.28) (30, 61.88) (31, 62.46) (32, 62.99) (33, 63.51) (34, 64.02) (35, 64.51) (36, 65.0) (37, 65.49) (38, 65.97) (39, 66.43) (40, 66.9) (41, 67.34) (42, 67.76) (43, 68.16) (44, 68.55) (45, 68.93) (46, 69.31) (47, 69.65) (48, 70.0) (49, 70.33) (50, 70.67) (51, 71.0) (52, 71.3) (53, 71.55) (54, 71.81) (55, 72.05) (56, 72.28) (57, 72.5) (58, 72.72) (59, 72.93) (60, 73.14) (61, 73.34) (62, 73.54) (63, 73.74) (64, 73.94) (65, 74.14) (66, 74.33) (67, 74.52) (68, 74.7) (69, 74.88) (70, 75.05) (71, 75.23) (72, 75.41) (73, 75.57) (74, 75.74) (75, 75.91) (76, 76.07) (77, 76.24) (78, 76.41) (79, 76.56) (80, 76.72) (81, 76.87) (82, 77.03) (83, 77.19) (84, 77.34) (85, 77.49) (86, 77.63) (87, 77.77) (88, 77.92) (89, 78.06) (90, 78.2) (91, 78.33) (92, 78.46) (93, 78.58) (94, 78.71) (95, 78.83) (96, 78.95) (97, 79.07) (98, 79.2) (99, 79.32) (100, 79.44) (101, 79.55) (102, 79.66) (103, 79.77) (104, 79.88) (105, 80.0) (106, 80.11) (107, 80.22) (108, 80.33) (109, 80.44) (110, 80.54) (111, 80.64) (112, 80.74) (113, 80.84) (114, 80.93) (115, 81.02) (116, 81.11) (117, 81.2) (118, 81.28) (119, 81.37) (120, 81.46) (121, 81.55) (122, 81.64) (123, 81.73) (124, 81.82) (125, 81.89) (126, 81.97) (127, 82.05) (128, 82.13) (129, 82.21) (130, 82.28) (131, 82.36) (132, 82.44) (133, 82.52) (134, 82.59) (135, 82.67) (136, 82.74) (137, 82.81) (138, 82.87) (139, 82.94) (140, 83.01) (141, 83.07) (142, 83.14) (143, 83.21) (144, 83.27) (145, 83.34) (146, 83.41) (147, 83.47) (148, 83.54) (149, 83.61) (150, 83.67) (151, 83.74) (152, 83.81) (153, 83.87) (154, 83.94) (155, 84.01) (156, 84.07) (157, 84.14) (158, 84.21) (159, 84.26) (160, 84.32) (161, 84.37) (162, 84.43) (163, 84.48) (164, 84.54) (165, 84.59) (166, 84.65) (167, 84.71) (168, 84.76) (169, 84.82) (170, 84.87) (171, 84.93) (172, 84.98) (173, 85.04) (174, 85.09) (175, 85.14) (176, 85.18) (177, 85.23) (178, 85.27) (179, 85.32) (180, 85.36) (181, 85.4) (182, 85.45) (183, 85.49) (184, 85.54) (185, 85.58) (186, 85.63) (187, 85.67) (188, 85.72) (189, 85.76) (190, 85.8) (191, 85.85) (192, 85.89) (193, 85.94) (194, 85.98) (195, 86.03) (196, 86.07) (197, 86.12) (198, 86.16) (199, 86.2) (200, 86.25) (201, 86.29) (202, 86.34) (203, 86.38) (204, 86.43) (205, 86.47) (206, 86.52) (207, 86.56) (208, 86.6) (209, 86.65) (210, 86.69) (211, 86.73) (212, 86.76) (213, 86.79) (214, 86.83) (215, 86.86) (216, 86.89) (217, 86.93) (218, 86.96) (219, 86.99) (220, 87.03) (221, 87.06) (222, 87.09) (223, 87.13) (224, 87.16) (225, 87.19) (226, 87.23) (227, 87.26) (228, 87.29) (229, 87.33) (230, 87.36) (231, 87.39) (232, 87.43) (233, 87.46) (234, 87.49) (235, 87.53) (236, 87.56) (237, 87.59) (238, 87.63) (239, 87.66) (240, 87.69) (241, 87.73) (242, 87.76) (243, 87.79) (244, 87.83) (245, 87.86) (246, 87.89) (247, 87.93) (248, 87.96) (249, 87.99) (250, 88.03) (251, 88.06) (252, 88.09) (253, 88.13) (254, 88.16) (255, 88.19) (256, 88.23) (257, 88.26) (258, 88.29) (259, 88.33) (260, 88.36) (261, 88.39) (262, 88.43) (263, 88.46) (264, 88.49) (265, 88.53) (266, 88.56) (267, 88.59) (268, 88.63) (269, 88.66) (270, 88.69) (271, 88.73) (272, 88.76) (273, 88.79) (274, 88.83) (275, 88.86) (276, 88.89) (277, 88.93) (278, 88.96) (279, 88.99) (280, 89.03) (281, 89.06) (282, 89.09) (283, 89.13) (284, 89.16) (285, 89.19) (286, 89.23) (287, 89.26) (288, 89.29) (289, 89.31) (290, 89.34) (291, 89.36) (292, 89.38) (293, 89.4) (294, 89.43) (295, 89.45) (296, 89.47) (297, 89.49) (298, 89.51) (299, 89.54) (300, 89.56) (301, 89.58) (302, 89.6) (303, 89.63) (304, 89.65) (305, 89.67) (306, 89.69) (307, 89.71) (308, 89.74) (309, 89.76) (310, 89.78) (311, 89.8) (312, 89.83) (313, 89.85) (314, 89.87) (315, 89.89) (316, 89.91) (317, 89.94) (318, 89.96) (319, 89.98) (320, 90.0) (321, 90.03) (322, 90.05) (323, 90.07) (324, 90.09) (325, 90.11) (326, 90.14) (327, 90.16) (328, 90.18) (329, 90.2) (330, 90.23) (331, 90.25) (332, 90.27) (333, 90.29) (334, 90.31) (335, 90.34) (336, 90.36) (337, 90.38) (338, 90.4) (339, 90.43) (340, 90.45) (341, 90.47) (342, 90.49) (343, 90.51) (344, 90.54) (345, 90.56) (346, 90.58) (347, 90.6) (348, 90.63) (349, 90.65) (350, 90.67) (351, 90.69) (352, 90.71) (353, 90.74) (354, 90.76) (355, 90.78) (356, 90.8) (357, 90.83) (358, 90.85) (359, 90.87) (360, 90.89) (361, 90.91) (362, 90.94) (363, 90.96) (364, 90.98) (365, 91.0) (366, 91.03) (367, 91.05) (368, 91.07) (369, 91.09) (370, 91.11) (371, 91.14) (372, 91.16) (373, 91.18) (374, 91.2) (375, 91.23) (376, 91.25) (377, 91.27) (378, 91.29) (379, 91.31) (380, 91.34) (381, 91.36) (382, 91.38) (383, 91.4) (384, 91.43) (385, 91.45) (386, 91.47) (387, 91.49) (388, 91.51) (389, 91.54) (390, 91.56) (391, 91.58) (392, 91.6) (393, 91.63) (394, 91.65) (395, 91.67) (396, 91.69) (397, 91.71) (398, 91.74) (399, 91.76) (400, 91.78) (401, 91.8) (402, 91.82) (403, 91.85) (404, 91.87) (405, 91.89) (406, 91.91) (407, 91.94) (408, 91.96) (409, 91.98) (410, 92.0) (411, 92.02) (412, 92.05) (413, 92.07) (414, 92.09) (415, 92.11) (416, 92.14) (417, 92.16) (418, 92.18) (419, 92.2) (420, 92.22) (421, 92.25) (422, 92.27) (423, 92.29) (424, 92.31) (425, 92.34) (426, 92.36) (427, 92.38) (428, 92.4) (429, 92.42) (430, 92.45) (431, 92.47) (432, 92.49) (433, 92.51) (434, 92.54) (435, 92.56) (436, 92.58) (437, 92.6) (438, 92.62) (439, 92.65) (440, 92.67) (441, 92.69) (442, 92.71) (443, 92.74) (444, 92.76) (445, 92.78) (446, 92.8) (447, 92.82) (448, 92.85) (449, 92.86) (450, 92.87) (451, 92.88) (452, 92.89) (453, 92.9) (454, 92.91) (455, 92.92) (456, 92.94) (457, 92.95) (458, 92.96) (459, 92.97) (460, 92.98) (461, 92.99) (462, 93.0) (463, 93.01) (464, 93.02) (465, 93.04) (466, 93.05) (467, 93.06) (468, 93.07) (469, 93.08) (470, 93.09) (471, 93.1) (472, 93.11) (473, 93.12) (474, 93.14) (475, 93.15) (476, 93.16) (477, 93.17) (478, 93.18) (479, 93.19) (480, 93.2) (481, 93.21) (482, 93.22) (483, 93.24) (484, 93.25) (485, 93.26) (486, 93.27) (487, 93.28) (488, 93.29) (489, 93.3) (490, 93.31) (491, 93.32) (492, 93.34) (493, 93.35) (494, 93.36) (495, 93.37) (496, 93.38) (497, 93.39) (498, 93.4) (499, 93.41) (500, 93.42) (501, 93.44) (502, 93.45) (503, 93.46) (504, 93.47) (505, 93.48) (506, 93.49) (507, 93.5) (508, 93.51) (509, 93.52) (510, 93.54) (511, 93.55) (512, 93.56) (513, 93.57) (514, 93.58) (515, 93.59) (516, 93.6) (517, 93.61) (518, 93.62) (519, 93.64) (520, 93.65) (521, 93.66) (522, 93.67) (523, 93.68) (524, 93.69) (525, 93.7) (526, 93.71) (527, 93.72) (528, 93.74) (529, 93.75) (530, 93.76) (531, 93.77) (532, 93.78) (533, 93.79) (534, 93.8) (535, 93.81) (536, 93.82) (537, 93.84) (538, 93.85) (539, 93.86) (540, 93.87) (541, 93.88) (542, 93.89) (543, 93.9) (544, 93.91) (545, 93.92) (546, 93.94) (547, 93.95) (548, 93.96) (549, 93.97) (550, 93.98) (551, 93.99) (552, 94.0) (553, 94.01) (554, 94.02) (555, 94.04) (556, 94.05) (557, 94.06) (558, 94.07) (559, 94.08) (560, 94.09) (561, 94.1) (562, 94.11) (563, 94.12) (564, 94.14) (565, 94.15) (566, 94.16) (567, 94.17) (568, 94.18) (569, 94.19) (570, 94.2) (571, 94.21) (572, 94.22) (573, 94.24) (574, 94.25) (575, 94.26) (576, 94.27) (577, 94.28) (578, 94.29) (579, 94.3) (580, 94.31) (581, 94.32) (582, 94.34) (583, 94.35) (584, 94.36) (585, 94.37) (586, 94.38) (587, 94.39) (588, 94.4) (589, 94.41) (590, 94.42) (591, 94.44) (592, 94.45) (593, 94.46) (594, 94.47) (595, 94.48) (596, 94.49) (597, 94.5) (598, 94.51) (599, 94.52) (600, 94.54) (601, 94.55) (602, 94.56) (603, 94.57) (604, 94.58) (605, 94.59) (606, 94.6) (607, 94.61) (608, 94.62) (609, 94.64) (610, 94.65) (611, 94.66) (612, 94.67) (613, 94.68) (614, 94.69) (615, 94.7) (616, 94.71) (617, 94.72) (618, 94.74) (619, 94.75) (620, 94.76) (621, 94.77) (622, 94.78) (623, 94.79) (624, 94.8) (625, 94.81) (626, 94.82) (627, 94.84) (628, 94.85) (629, 94.86) (630, 94.87) (631, 94.88) (632, 94.89) (633, 94.9) (634, 94.91) (635, 94.92) (636, 94.94) (637, 94.95) (638, 94.96) (639, 94.97) (640, 94.98) (641, 94.99) (642, 95.0) (643, 95.01) (644, 95.02) (645, 95.03) (646, 95.05) (647, 95.06) (648, 95.07) (649, 95.08) (650, 95.09) (651, 95.1) (652, 95.11) (653, 95.12) (654, 95.13) (655, 95.15) (656, 95.16) (657, 95.17) (658, 95.18) (659, 95.19) (660, 95.2) (661, 95.21) (662, 95.22) (663, 95.23) (664, 95.25) (665, 95.26) (666, 95.27) (667, 95.28) (668, 95.29) (669, 95.3) (670, 95.31) (671, 95.32) (672, 95.33) (673, 95.35) (674, 95.36) (675, 95.37) (676, 95.38) (677, 95.39) (678, 95.4) (679, 95.41) (680, 95.42) (681, 95.43) (682, 95.45) (683, 95.46) (684, 95.47) (685, 95.48) (686, 95.49) (687, 95.5) (688, 95.51) (689, 95.52) (690, 95.53) (691, 95.55) (692, 95.56) (693, 95.57) (694, 95.58) (695, 95.59) (696, 95.6) (697, 95.61) (698, 95.62) (699, 95.63) (700, 95.65) (701, 95.66) (702, 95.67) (703, 95.68) (704, 95.69) (705, 95.7) (706, 95.71) (707, 95.72) (708, 95.73) (709, 95.75) (710, 95.76) (711, 95.77) (712, 95.78) (713, 95.79) (714, 95.8) (715, 95.81) (716, 95.82) (717, 95.83) (718, 95.85) (719, 95.86) (720, 95.87) (721, 95.88) (722, 95.89) (723, 95.9) (724, 95.91) (725, 95.92) (726, 95.93) (727, 95.95) (728, 95.96) (729, 95.97) (730, 95.98) (731, 95.99) (732, 96.0) (733, 96.01) (734, 96.02) (735, 96.03) (736, 96.05) (737, 96.06) (738, 96.07) (739, 96.08) (740, 96.09) (741, 96.1) (742, 96.11) (743, 96.12) (744, 96.13) (745, 96.15) (746, 96.16) (747, 96.17) (748, 96.18) (749, 96.19) (750, 96.2) (751, 96.21) (752, 96.22) (753, 96.23) (754, 96.25) (755, 96.26) (756, 96.27) (757, 96.28) (758, 96.29) (759, 96.3) (760, 96.31) (761, 96.32) (762, 96.33) (763, 96.35) (764, 96.36) (765, 96.37) (766, 96.38) (767, 96.39) (768, 96.4) (769, 96.41) (770, 96.42) (771, 96.43) (772, 96.45) (773, 96.46) (774, 96.47) (775, 96.48) (776, 96.49) (777, 96.5) (778, 96.51) (779, 96.52) (780, 96.53) (781, 96.55) (782, 96.56) (783, 96.57) (784, 96.58) (785, 96.59) (786, 96.6) (787, 96.61) (788, 96.62) (789, 96.63) (790, 96.65) (791, 96.66) (792, 96.67) (793, 96.68) (794, 96.69) (795, 96.7) (796, 96.71) (797, 96.72) (798, 96.73) (799, 96.75) (800, 96.76) (801, 96.77) (802, 96.78) (803, 96.79) (804, 96.8) (805, 96.81) (806, 96.82) (807, 96.83) (808, 96.85) (809, 96.86) (810, 96.87) (811, 96.88) (812, 96.89) (813, 96.9) (814, 96.91) (815, 96.92) (816, 96.93) (817, 96.95) (818, 96.96) (819, 96.97) (820, 96.98) (821, 96.99) (822, 97.0) (823, 97.01) (824, 97.02) (825, 97.03) (826, 97.05) (827, 97.06) (828, 97.07) (829, 97.08) (830, 97.09) (831, 97.1) (832, 97.11) (833, 97.12) (834, 97.13) (835, 97.15) (836, 97.16) (837, 97.17) (838, 97.18) (839, 97.19) (840, 97.2) (841, 97.21) (842, 97.22) (843, 97.23) (844, 97.25) (845, 97.26) (846, 97.27) (847, 97.28) (848, 97.29) (849, 97.3) (850, 97.31) (851, 97.32) (852, 97.33) (853, 97.35) (854, 97.36) (855, 97.37) (856, 97.38) (857, 97.39) (858, 97.4) (859, 97.41) (860, 97.42) (861, 97.43) (862, 97.45) (863, 97.46) (864, 97.47) (865, 97.48) (866, 97.49) (867, 97.5) (868, 97.51) (869, 97.52) (870, 97.53) (871, 97.55) (872, 97.56) (873, 97.57) (874, 97.58) (875, 97.59) (876, 97.6) (877, 97.61) (878, 97.62) (879, 97.63) (880, 97.65) (881, 97.66) (882, 97.67) (883, 97.68) (884, 97.69) (885, 97.7) (886, 97.71) (887, 97.72) (888, 97.73) (889, 97.75) (890, 97.76) (891, 97.77) (892, 97.78) (893, 97.79) (894, 97.8) (895, 97.81) (896, 97.82) (897, 97.83) (898, 97.85) (899, 97.86) (900, 97.87) (901, 97.88) (902, 97.89) (903, 97.9) (904, 97.91) (905, 97.92) (906, 97.93) (907, 97.95) (908, 97.96) (909, 97.97) (910, 97.98) (911, 97.99) (912, 98.0) (913, 98.01) (914, 98.02) (915, 98.03) (916, 98.05) (917, 98.06) (918, 98.07) (919, 98.08) (920, 98.09) (921, 98.1) (922, 98.11) (923, 98.12) (924, 98.13) (925, 98.15) (926, 98.16) (927, 98.17) (928, 98.18) (929, 98.19) (930, 98.2) (931, 98.21) (932, 98.22) (933, 98.23) (934, 98.25) (935, 98.26) (936, 98.27) (937, 98.28) (938, 98.29) (939, 98.3) (940, 98.31) (941, 98.32) (942, 98.33) (943, 98.34) (944, 98.36) (945, 98.37) (946, 98.38) (947, 98.39) (948, 98.4) (949, 98.41) (950, 98.42) (951, 98.43) (952, 98.44) (953, 98.46) (954, 98.47) (955, 98.48) (956, 98.49) (957, 98.5) (958, 98.51) (959, 98.52) (960, 98.53) (961, 98.54) (962, 98.56) (963, 98.57) (964, 98.58) (965, 98.59) (966, 98.6) (967, 98.61) (968, 98.62) (969, 98.63) (970, 98.64) (971, 98.66) (972, 98.67) (973, 98.68) (974, 98.69) (975, 98.7) (976, 98.71) (977, 98.72) (978, 98.73) (979, 98.74) (980, 98.76) (981, 98.77) (982, 98.78) (983, 98.79) (984, 98.8) (985, 98.81) (986, 98.82) (987, 98.83) (988, 98.84) (989, 98.86) (990, 98.87) (991, 98.88) (992, 98.89) (993, 98.9) (994, 98.91) (995, 98.92) (996, 98.93) (997, 98.94) (998, 98.96) (999, 98.97) (1000, 98.98) (1001, 98.99) (1002, 99.0) (1003, 99.01) (1004, 99.02) (1005, 99.03) (1006, 99.04) (1007, 99.06) (1008, 99.07) (1009, 99.08) (1010, 99.09) (1011, 99.1) (1012, 99.11) (1013, 99.12) (1014, 99.13) (1015, 99.14) (1016, 99.16) (1017, 99.17) (1018, 99.18) (1019, 99.19) (1020, 99.2) (1021, 99.21) (1022, 99.22) (1023, 99.23) (1024, 99.24) (1025, 99.26) (1026, 99.27) (1027, 99.28) (1028, 99.29) (1029, 99.3) (1030, 99.31) (1031, 99.32) (1032, 99.33) (1033, 99.34) (1034, 99.36) (1035, 99.37) (1036, 99.38) (1037, 99.39) (1038, 99.4) (1039, 99.41) (1040, 99.42) (1041, 99.43) (1042, 99.44) (1043, 99.46) (1044, 99.47) (1045, 99.48) (1046, 99.49) (1047, 99.5) (1048, 99.51) (1049, 99.52) (1050, 99.53) (1051, 99.54) (1052, 99.56) (1053, 99.57) (1054, 99.58) (1055, 99.59) (1056, 99.6) (1057, 99.61) (1058, 99.62) (1059, 99.63) (1060, 99.64) (1061, 99.66) (1062, 99.67) (1063, 99.68) (1064, 99.69) (1065, 99.7) (1066, 99.71) (1067, 99.72) (1068, 99.73) (1069, 99.74) (1070, 99.76) (1071, 99.77) (1072, 99.78) (1073, 99.79) (1074, 99.8) (1075, 99.81) (1076, 99.82) (1077, 99.83) (1078, 99.84) (1079, 99.86) (1080, 99.87) (1081, 99.88) (1082, 99.89) (1083, 99.9) (1084, 99.91) (1085, 99.92) (1086, 99.93) (1087, 99.94) (1088, 99.96) (1089, 99.97) (1090, 99.98) (1091, 99.99) (1092, 100.0) };
			\addlegendentry{\strut .nl}
			
			\addplot[blue, thick] 
coordinates { (1, 33.45) (2, 51.98) (3, 56.8) (4, 61.55) (5, 65.1) (6, 68.13) (7, 70.59) (8, 73.0) (9, 74.8) (10, 76.24) (11, 77.59) (12, 78.92) (13, 80.17) (14, 81.39) (15, 82.45) (16, 83.43) (17, 84.15) (18, 84.86) (19, 85.49) (20, 86.13) (21, 86.74) (22, 87.31) (23, 87.8) (24, 88.21) (25, 88.6) (26, 88.98) (27, 89.35) (28, 89.71) (29, 90.06) (30, 90.38) (31, 90.71) (32, 91.01) (33, 91.32) (34, 91.6) (35, 91.88) (36, 92.12) (37, 92.35) (38, 92.57) (39, 92.77) (40, 92.97) (41, 93.17) (42, 93.36) (43, 93.55) (44, 93.73) (45, 93.91) (46, 94.09) (47, 94.26) (48, 94.42) (49, 94.57) (50, 94.73) (51, 94.88) (52, 95.03) (53, 95.16) (54, 95.29) (55, 95.42) (56, 95.54) (57, 95.66) (58, 95.77) (59, 95.86) (60, 95.95) (61, 96.05) (62, 96.14) (63, 96.23) (64, 96.31) (65, 96.4) (66, 96.48) (67, 96.56) (68, 96.63) (69, 96.7) (70, 96.78) (71, 96.85) (72, 96.91) (73, 96.98) (74, 97.04) (75, 97.1) (76, 97.17) (77, 97.23) (78, 97.29) (79, 97.36) (80, 97.41) (81, 97.46) (82, 97.51) (83, 97.57) (84, 97.62) (85, 97.67) (86, 97.72) (87, 97.78) (88, 97.83) (89, 97.88) (90, 97.92) (91, 97.97) (92, 98.01) (93, 98.05) (94, 98.09) (95, 98.14) (96, 98.18) (97, 98.22) (98, 98.26) (99, 98.3) (100, 98.34) (101, 98.37) (102, 98.4) (103, 98.43) (104, 98.46) (105, 98.49) (106, 98.53) (107, 98.56) (108, 98.59) (109, 98.62) (110, 98.65) (111, 98.68) (112, 98.71) (113, 98.75) (114, 98.78) (115, 98.81) (116, 98.84) (117, 98.87) (118, 98.9) (119, 98.94) (120, 98.96) (121, 98.98) (122, 99.0) (123, 99.02) (124, 99.04) (125, 99.06) (126, 99.08) (127, 99.1) (128, 99.13) (129, 99.15) (130, 99.17) (131, 99.19) (132, 99.21) (133, 99.23) (134, 99.25) (135, 99.27) (136, 99.29) (137, 99.32) (138, 99.34) (139, 99.36) (140, 99.38) (141, 99.4) (142, 99.42) (143, 99.44) (144, 99.46) (145, 99.48) (146, 99.5) (147, 99.53) (148, 99.55) (149, 99.57) (150, 99.59) (151, 99.6) (152, 99.61) (153, 99.62) (154, 99.63) (155, 99.64) (156, 99.65) (157, 99.66) (158, 99.67) (159, 99.68) (160, 99.69) (161, 99.71) (162, 99.72) (163, 99.73) (164, 99.74) (165, 99.75) (166, 99.76) (167, 99.77) (168, 99.78) (169, 99.79) (170, 99.8) (171, 99.81) (172, 99.82) (173, 99.83) (174, 99.84) (175, 99.85) (176, 99.86) (177, 99.87) (178, 99.88) (179, 99.89) (180, 99.91) (181, 99.92) (182, 99.93) (183, 99.94) (184, 99.95) (185, 99.96) (186, 99.97) (187, 99.98) (188, 99.99) (189, 100.0) };
			\addlegendentry{\strut .fr}
			
			\addplot[green!60!black, thick] 
coordinates { (1, 35.25) (2, 43.98) (3, 48.81) (4, 53.25) (5, 57.1) (6, 60.41) (7, 63.12) (8, 65.67) (9, 67.84) (10, 69.85) (11, 71.8) (12, 73.64) (13, 75.49) (14, 77.33) (15, 78.69) (16, 79.88) (17, 81.02) (18, 81.89) (19, 82.65) (20, 83.41) (21, 84.11) (22, 84.65) (23, 85.2) (24, 85.74) (25, 86.28) (26, 86.77) (27, 87.2) (28, 87.64) (29, 88.07) (30, 88.5) (31, 88.94) (32, 89.32) (33, 89.64) (34, 89.97) (35, 90.29) (36, 90.62) (37, 90.89) (38, 91.16) (39, 91.43) (40, 91.7) (41, 91.97) (42, 92.25) (43, 92.52) (44, 92.79) (45, 93.0) (46, 93.22) (47, 93.44) (48, 93.66) (49, 93.87) (50, 94.09) (51, 94.25) (52, 94.41) (53, 94.58) (54, 94.74) (55, 94.9) (56, 95.07) (57, 95.17) (58, 95.28) (59, 95.39) (60, 95.5) (61, 95.61) (62, 95.72) (63, 95.82) (64, 95.93) (65, 96.04) (66, 96.15) (67, 96.26) (68, 96.37) (69, 96.48) (70, 96.58) (71, 96.69) (72, 96.8) (73, 96.91) (74, 97.02) (75, 97.13) (76, 97.23) (77, 97.34) (78, 97.4) (79, 97.45) (80, 97.51) (81, 97.56) (82, 97.61) (83, 97.67) (84, 97.72) (85, 97.78) (86, 97.83) (87, 97.89) (88, 97.94) (89, 97.99) (90, 98.05) (91, 98.1) (92, 98.16) (93, 98.21) (94, 98.26) (95, 98.32) (96, 98.37) (97, 98.43) (98, 98.48) (99, 98.54) (100, 98.59) (101, 98.64) (102, 98.7) (103, 98.75) (104, 98.81) (105, 98.86) (106, 98.92) (107, 98.97) (108, 99.02) (109, 99.08) (110, 99.13) (111, 99.19) (112, 99.24) (113, 99.3) (114, 99.35) (115, 99.4) (116, 99.46) (117, 99.51) (118, 99.57) (119, 99.62) (120, 99.67) (121, 99.73) (122, 99.78) (123, 99.84) (124, 99.89) (125, 99.95) (126, 100.0) };
			\addlegendentry{\strut .no}
			
			\addplot[orange, thick] 
coordinates { (1, 12.26) (2, 20.8) (3, 28.1) (4, 34.85) (5, 40.17) (6, 45.15) (7, 49.44) (8, 53.11) (9, 56.17) (10, 59.0) (11, 60.8) (12, 62.59) (13, 63.94) (14, 65.27) (15, 66.48) (16, 67.66) (17, 68.84) (18, 69.73) (19, 70.55) (20, 71.33) (21, 72.01) (22, 72.65) (23, 73.27) (24, 73.9) (25, 74.49) (26, 75.07) (27, 75.57) (28, 76.05) (29, 76.51) (30, 76.96) (31, 77.38) (32, 77.77) (33, 78.15) (34, 78.52) (35, 78.88) (36, 79.22) (37, 79.55) (38, 79.87) (39, 80.18) (40, 80.48) (41, 80.75) (42, 81.01) (43, 81.27) (44, 81.52) (45, 81.77) (46, 82.01) (47, 82.24) (48, 82.47) (49, 82.69) (50, 82.9) (51, 83.11) (52, 83.32) (53, 83.52) (54, 83.71) (55, 83.9) (56, 84.07) (57, 84.24) (58, 84.41) (59, 84.58) (60, 84.74) (61, 84.9) (62, 85.07) (63, 85.22) (64, 85.38) (65, 85.53) (66, 85.69) (67, 85.83) (68, 85.98) (69, 86.11) (70, 86.25) (71, 86.39) (72, 86.52) (73, 86.65) (74, 86.78) (75, 86.91) (76, 87.03) (77, 87.16) (78, 87.29) (79, 87.41) (80, 87.54) (81, 87.66) (82, 87.78) (83, 87.9) (84, 88.01) (85, 88.13) (86, 88.24) (87, 88.35) (88, 88.45) (89, 88.55) (90, 88.65) (91, 88.75) (92, 88.85) (93, 88.95) (94, 89.05) (95, 89.15) (96, 89.24) (97, 89.33) (98, 89.42) (99, 89.5) (100, 89.58) (101, 89.66) (102, 89.74) (103, 89.82) (104, 89.89) (105, 89.96) (106, 90.03) (107, 90.11) (108, 90.18) (109, 90.25) (110, 90.32) (111, 90.39) (112, 90.45) (113, 90.52) (114, 90.58) (115, 90.64) (116, 90.71) (117, 90.77) (118, 90.83) (119, 90.9) (120, 90.95) (121, 91.01) (122, 91.06) (123, 91.12) (124, 91.17) (125, 91.22) (126, 91.28) (127, 91.33) (128, 91.39) (129, 91.44) (130, 91.5) (131, 91.55) (132, 91.61) (133, 91.66) (134, 91.72) (135, 91.76) (136, 91.81) (137, 91.85) (138, 91.9) (139, 91.94) (140, 91.99) (141, 92.03) (142, 92.08) (143, 92.13) (144, 92.17) (145, 92.22) (146, 92.26) (147, 92.31) (148, 92.35) (149, 92.4) (150, 92.44) (151, 92.49) (152, 92.53) (153, 92.58) (154, 92.63) (155, 92.67) (156, 92.72) (157, 92.76) (158, 92.81) (159, 92.84) (160, 92.88) (161, 92.92) (162, 92.95) (163, 92.99) (164, 93.03) (165, 93.06) (166, 93.1) (167, 93.13) (168, 93.17) (169, 93.21) (170, 93.24) (171, 93.28) (172, 93.32) (173, 93.35) (174, 93.39) (175, 93.43) (176, 93.46) (177, 93.5) (178, 93.53) (179, 93.57) (180, 93.61) (181, 93.64) (182, 93.68) (183, 93.72) (184, 93.74) (185, 93.77) (186, 93.8) (187, 93.83) (188, 93.85) (189, 93.88) (190, 93.91) (191, 93.93) (192, 93.96) (193, 93.99) (194, 94.02) (195, 94.04) (196, 94.07) (197, 94.1) (198, 94.13) (199, 94.15) (200, 94.18) (201, 94.21) (202, 94.23) (203, 94.26) (204, 94.29) (205, 94.32) (206, 94.34) (207, 94.37) (208, 94.4) (209, 94.43) (210, 94.45) (211, 94.48) (212, 94.51) (213, 94.53) (214, 94.56) (215, 94.59) (216, 94.62) (217, 94.64) (218, 94.67) (219, 94.7) (220, 94.73) (221, 94.75) (222, 94.78) (223, 94.81) (224, 94.83) (225, 94.86) (226, 94.89) (227, 94.92) (228, 94.94) (229, 94.97) (230, 95.0) (231, 95.03) (232, 95.05) (233, 95.08) (234, 95.11) (235, 95.14) (236, 95.16) (237, 95.19) (238, 95.22) (239, 95.24) (240, 95.26) (241, 95.28) (242, 95.3) (243, 95.32) (244, 95.34) (245, 95.35) (246, 95.37) (247, 95.39) (248, 95.41) (249, 95.43) (250, 95.44) (251, 95.46) (252, 95.48) (253, 95.5) (254, 95.52) (255, 95.54) (256, 95.55) (257, 95.57) (258, 95.59) (259, 95.61) (260, 95.63) (261, 95.64) (262, 95.66) (263, 95.68) (264, 95.7) (265, 95.72) (266, 95.74) (267, 95.75) (268, 95.77) (269, 95.79) (270, 95.81) (271, 95.83) (272, 95.84) (273, 95.86) (274, 95.88) (275, 95.9) (276, 95.92) (277, 95.94) (278, 95.95) (279, 95.97) (280, 95.99) (281, 96.01) (282, 96.03) (283, 96.04) (284, 96.06) (285, 96.08) (286, 96.1) (287, 96.12) (288, 96.14) (289, 96.15) (290, 96.17) (291, 96.19) (292, 96.21) (293, 96.23) (294, 96.24) (295, 96.26) (296, 96.28) (297, 96.3) (298, 96.32) (299, 96.34) (300, 96.35) (301, 96.37) (302, 96.39) (303, 96.41) (304, 96.43) (305, 96.44) (306, 96.46) (307, 96.48) (308, 96.5) (309, 96.52) (310, 96.54) (311, 96.55) (312, 96.57) (313, 96.59) (314, 96.61) (315, 96.63) (316, 96.64) (317, 96.66) (318, 96.68) (319, 96.7) (320, 96.72) (321, 96.74) (322, 96.75) (323, 96.77) (324, 96.79) (325, 96.81) (326, 96.83) (327, 96.84) (328, 96.86) (329, 96.88) (330, 96.9) (331, 96.92) (332, 96.94) (333, 96.95) (334, 96.97) (335, 96.99) (336, 97.01) (337, 97.03) (338, 97.04) (339, 97.06) (340, 97.08) (341, 97.1) (342, 97.12) (343, 97.14) (344, 97.15) (345, 97.17) (346, 97.18) (347, 97.19) (348, 97.2) (349, 97.21) (350, 97.22) (351, 97.23) (352, 97.24) (353, 97.24) (354, 97.25) (355, 97.26) (356, 97.27) (357, 97.28) (358, 97.29) (359, 97.3) (360, 97.31) (361, 97.32) (362, 97.33) (363, 97.34) (364, 97.34) (365, 97.35) (366, 97.36) (367, 97.37) (368, 97.38) (369, 97.39) (370, 97.4) (371, 97.41) (372, 97.42) (373, 97.43) (374, 97.44) (375, 97.44) (376, 97.45) (377, 97.46) (378, 97.47) (379, 97.48) (380, 97.49) (381, 97.5) (382, 97.51) (383, 97.52) (384, 97.53) (385, 97.54) (386, 97.54) (387, 97.55) (388, 97.56) (389, 97.57) (390, 97.58) (391, 97.59) (392, 97.6) (393, 97.61) (394, 97.62) (395, 97.63) (396, 97.64) (397, 97.64) (398, 97.65) (399, 97.66) (400, 97.67) (401, 97.68) (402, 97.69) (403, 97.7) (404, 97.71) (405, 97.72) (406, 97.73) (407, 97.74) (408, 97.74) (409, 97.75) (410, 97.76) (411, 97.77) (412, 97.78) (413, 97.79) (414, 97.8) (415, 97.81) (416, 97.82) (417, 97.83) (418, 97.84) (419, 97.84) (420, 97.85) (421, 97.86) (422, 97.87) (423, 97.88) (424, 97.89) (425, 97.9) (426, 97.91) (427, 97.92) (428, 97.93) (429, 97.94) (430, 97.94) (431, 97.95) (432, 97.96) (433, 97.97) (434, 97.98) (435, 97.99) (436, 98.0) (437, 98.01) (438, 98.02) (439, 98.03) (440, 98.04) (441, 98.04) (442, 98.05) (443, 98.06) (444, 98.07) (445, 98.08) (446, 98.09) (447, 98.1) (448, 98.11) (449, 98.12) (450, 98.13) (451, 98.14) (452, 98.14) (453, 98.15) (454, 98.16) (455, 98.17) (456, 98.18) (457, 98.19) (458, 98.2) (459, 98.21) (460, 98.22) (461, 98.23) (462, 98.24) (463, 98.24) (464, 98.25) (465, 98.26) (466, 98.27) (467, 98.28) (468, 98.29) (469, 98.3) (470, 98.31) (471, 98.32) (472, 98.33) (473, 98.34) (474, 98.35) (475, 98.35) (476, 98.36) (477, 98.37) (478, 98.38) (479, 98.39) (480, 98.4) (481, 98.41) (482, 98.42) (483, 98.43) (484, 98.44) (485, 98.45) (486, 98.45) (487, 98.46) (488, 98.47) (489, 98.48) (490, 98.49) (491, 98.5) (492, 98.51) (493, 98.52) (494, 98.53) (495, 98.54) (496, 98.55) (497, 98.55) (498, 98.56) (499, 98.57) (500, 98.58) (501, 98.59) (502, 98.6) (503, 98.61) (504, 98.62) (505, 98.63) (506, 98.64) (507, 98.65) (508, 98.65) (509, 98.66) (510, 98.67) (511, 98.68) (512, 98.69) (513, 98.7) (514, 98.71) (515, 98.72) (516, 98.73) (517, 98.74) (518, 98.75) (519, 98.75) (520, 98.76) (521, 98.77) (522, 98.78) (523, 98.79) (524, 98.8) (525, 98.81) (526, 98.82) (527, 98.83) (528, 98.84) (529, 98.85) (530, 98.85) (531, 98.86) (532, 98.87) (533, 98.88) (534, 98.89) (535, 98.9) (536, 98.91) (537, 98.92) (538, 98.93) (539, 98.94) (540, 98.95) (541, 98.95) (542, 98.96) (543, 98.97) (544, 98.98) (545, 98.99) (546, 99.0) (547, 99.01) (548, 99.02) (549, 99.03) (550, 99.04) (551, 99.05) (552, 99.05) (553, 99.06) (554, 99.07) (555, 99.08) (556, 99.09) (557, 99.1) (558, 99.11) (559, 99.12) (560, 99.13) (561, 99.14) (562, 99.15) (563, 99.15) (564, 99.16) (565, 99.17) (566, 99.18) (567, 99.19) (568, 99.2) (569, 99.21) (570, 99.22) (571, 99.23) (572, 99.24) (573, 99.25) (574, 99.25) (575, 99.26) (576, 99.27) (577, 99.28) (578, 99.29) (579, 99.3) (580, 99.31) (581, 99.32) (582, 99.33) (583, 99.34) (584, 99.35) (585, 99.35) (586, 99.36) (587, 99.37) (588, 99.38) (589, 99.39) (590, 99.4) (591, 99.41) (592, 99.42) (593, 99.43) (594, 99.44) (595, 99.45) (596, 99.45) (597, 99.46) (598, 99.47) (599, 99.48) (600, 99.49) (601, 99.5) (602, 99.51) (603, 99.52) (604, 99.53) (605, 99.54) (606, 99.55) (607, 99.55) (608, 99.56) (609, 99.57) (610, 99.58) (611, 99.59) (612, 99.6) (613, 99.61) (614, 99.62) (615, 99.63) (616, 99.64) (617, 99.65) (618, 99.65) (619, 99.66) (620, 99.67) (621, 99.68) (622, 99.69) (623, 99.7) (624, 99.71) (625, 99.72) (626, 99.73) (627, 99.74) (628, 99.75) (629, 99.75) (630, 99.76) (631, 99.77) (632, 99.78) (633, 99.79) (634, 99.8) (635, 99.81) (636, 99.82) (637, 99.83) (638, 99.84) (639, 99.85) (640, 99.85) (641, 99.86) (642, 99.87) (643, 99.88) (644, 99.89) (645, 99.9) (646, 99.91) (647, 99.92) (648, 99.93) (649, 99.94) (650, 99.95) (651, 99.95) (652, 99.96) (653, 99.97) (654, 99.98) (655, 99.99) (656, 100.0) };
			\addlegendentry{\strut .uk}
			
		\end{axis}
	\end{tikzpicture}
	}
\end{center}
\caption{The market share for the TLDs .nl, .fr, .no, and .uk, shown as a cumulative percentage for the top N registrars.}
\label{fig:market-share}
\end{figure}

We also used the same Cloudflare Radar data used in Figure~\ref{fig:top-registrars} for .fr, .no, and .uk. We show their agent market share in Figure~\ref{fig:market-share}. We picked these three because these ccTLDs are mainly used by things related to France, Norway, and the United Kingdom respectively, all three countries are also considered role models by the ITU, and they also all have RDAP availability that includes agent information (as WHOIS data can be tricky to parse \cite{liu_2015}) so we could analyse their data. For all three we see a large local presence in registrars, although American-based ones are popular in the United Kingdom as well. We also noticed some overlap, where \agent{C}, \agent{CSC Global}, and \agent{MarkMonitor} tend to be popular at these three ccTLDs too. 

The impact of a domain takeover we describe in Section~\ref{sec:impact} is not specific to any country or (cc)TLD. Only the likelihood of occurrence changes. We believe the risk in other role model countries is similar, and might be higher in other countries, for example in countries where TOTP is not as widespread.

\section{Disclosure Process}\label{sec:disclosure}
We decided to contact the agents from Figure~\ref{fig:totp-protection} that did not apply per-account rate limiting. \agent{A}, \agent{C}, \agent{E}, \agent{I}, and \agent{J} have a \textit{security.txt} \cite{rfc9116} set up. \agent{B}, \agent{F}, \agent{G} do not have a security.txt page, but do have a responsible disclosure page set up. Only \agent{D} and \agent{H} have neither. 

We contacted \agent{C}, \agent{E}, and \agent{J} using their contact address in security.txt, \agent{B}, \agent{F}, and \agent{G} via their responsible disclosure pages, and \agent{D} and \agent{H} and via their general contact form. We sent them a message along the following lines (slightly adjusted for those who did do IP rate limiting):

\begin{quote}
	\textit{Subject: TOTP rate limiting}
	
	Dear <organisation>,
	
	At
	the University of Twente 
	, we have been conducting research into the security of domain names in the Netherlands and how registrars and resellers prevent unauthorised access to, for example, the web portal. Part of this involves analysing which security methods are used to prevent brute force attacks on two-factor authentication. 
	
	We noticed that it appears that no rate limiting is being applied, which is why we are sending this email.
	
	Our test setup is as follows: we log in normally with a username and password, then we try 100 incorrect TOTP codes in ~1 second, after which we enter the correct TOTP code. We analyse whether and how rate limiting is applied based on the responses we receive from the 100 incorrect TOTP codes. 
	
	RFC 4226 section 7.3 stipulates that rate limiting must be implemented per user and not per session, in order to counter parallel attacks. We have limited ourselves to 100 attempts so as not to put unnecessary pressure on your servers. Without rate limiting, with 100 requests per second, a correct TOTP code can be guessed in just under an hour on average. 
	
	Is our assumption that no rate limiting is applied correct? And if so, are there plans to implement rate limiting in the next three months? Our goal is not to `name and shame'; we therefore do not publish any names in the study. Nevertheless, we would like to mention that parties have implemented rate limiting, so we would appreciate a response.
	
	We look forward to hearing from you. Please feel free to contact us with any questions or comments.
	
	Yours faithfully,\\
	<name>
	
	<email>\\
	<phone number>
	
	<address>
\end{quote}

We received a confirmation within a day from two out of eight notifications, namely from \agent{D} and \agent{H}, and also from \agent{E} after a week. We have not received a message from the others till this date. None have stated that the issue was resolved.


\section{Discussion}\label{sec:discussion}
Signing up for the agents listed in Figure~\ref{fig:registrations} was more difficult than expected. Two required manual intervention by customer support after we contacted them that the registration had seemingly not worked. One kept the old DS-record (a record used by DNSSEC) when we moved to our own DNS name servers. When we asked them how to remove or change that, they disabled DNSSEC for us without requiring any verification. 

However, as we show in Section~\ref{sec:phone-calls}, when trying to gain access to an account, all agents seem to have procedures in place to prevent unauthorised takeovers. We did not expect this result. We also expected the registrar to ask verification question such as email, telephone number, address, and date of birth, but none of them requested that information.

These events, the detected protection methods against TOTP brute forcing, as well as the general impression of the online portals, gave us the feeling that many of these systems are rather brittle, but the limitations known to the organisations. Sadly we have no way to fully verify that without access to the source code.

We were surprised to find that none of the agents sent us an email after a failed TOTP login attempt. Whilst the result from Figure~\ref{fig:totp-protection} do show that some of them prevent against brute forcing, it seems like TOTP support and the security implications were not fully thought through.

We were also surprised to find that there are simultaneously several popular agents such as \agent{CSC Global} and \agent{MarkMonitor} that offer brand protection services, indicating that there is a realisation at organisations that their domain name is important, yet simultaneously a lack of formal impact models of what would happen if a domain name were to be taken over.

\subsection{Limitations}
There are some limitations to our approach, which we want to specifically highlight separately. 

As alluded to in Section~\ref{sec:determining-agents}, we could not test corporate registrars, which are the most prominent in the 480 most popular domain names as we describe in Section~\ref{sec:most-popular-domains}. Additionally, even though we believe our results are applicable outside the Netherlands as well as we describe in Section~\ref{sec:countries}, care and attention is still required when applying these results outside the Netherlands. Furthermore, it should be noted that country-code TLDs such as .nl can also be used outside the context of the Netherlands and vice versa.

The password availability metrics mentioned in Section~\ref{sec:password-availability} assumes the same password reuse numbers as found in other research. These numbers may also vary wildly, e.g., in case there is a recent large-scale password breach. Ethical and legal limitations prevent us from verifying these claims. Similar ethical and legal limitations prevent us from sending a fake ID card scan as was requested in some of the helpdesk calls from Section~\ref{sec:phone-calls}.

\section{Recommendations}\label{sec:recommendations}
Whilst we believe that it is worthwhile to analyse the market as a whole, we also want to share a concrete list of points to consider for organisations and individuals:

\begin{enumerate}
	\item Make an impact assessment of what would happen if your domain name is taken over. In Section~\ref{sec:impact} we describe some of the potential impacts, but as every organisation is different, this list will be different as well;
	\item Monitor your domain name, including its DNS records, DNSSEC configuration, and WHOIS data. Changes here can be an indication that someone is tampering with your domain;
	\item Pick a registrar wisely. There are still registrars out there that do not support things like second factor authentication. Having a trustworthy and secure registrar is thus crucial;
	\item Use an email alias for your contact address in the WHOIS, and do not use this email to sign up at other places or for logging in to the administrative portal.
\end{enumerate}

\section{Conclusion}\label{sec:conclusion}
We analysed the risk of a domain name takeover by examining the technical likelihood of a domain takeover and the impact if it were to occur.

Firstly, we looked at the security of domain names in the Netherlands. When it comes to the security at popular domain registrars in the Netherlands. The positive part is that multi-factor authentication is available in all cases we examined and processes are in place to verify the identity of the domain holder, even over the phone. Additionally, larger organisations seem to use their own registrar or a registrar specifically aimed at protecting brands. However, things as TOTP are still often easy to brute force in most cases, and notifications and confirmation of account access and changes are often lacking, making it potentially difficult to notice when an account has been breached. Email addresses for domain management are reused for other purposes, and are frequently personal email addresses that appear in public breaches and data leaks.

Secondly, we find that gaining illicit access to a domain name can have great impact. We analysed the impact of a domain name takeover, and find that its potential consequences are significant, especially with the uptake in use of cloud services, and likely between that of a DDoS attack and a ransomware attack.

These two factors combined lead us to believe the risk of a domain name is potentially great. We have no reason to believe these results are unique to The Netherlands, and are likely relevant throughout Europe and even globally. We therefore urge organisations to take the impact of a domain takeover for their organisation into consideration, and provide recommendations and points for organisations to consider.

\subsection*{Future Work}\label{sec:future-work}
We believe it is worthwhile to repeat this experiment in the future, especially after the disclosure we describe in Section~\ref{sec:disclosure}. Additionally, we believe it is worth investigating this with different markets in and outside Europe to confirm our assumptions from Section~\ref{sec:countries}. 



\bibliographystyle{plainurl}
\bibliography{paper}


\clearpage
\begin{figure*}[hbt!]
	\appendix
	\section{Phone flow chart}\label{sec:flowchart}
	\includegraphics[height=20cm, keepaspectratio]{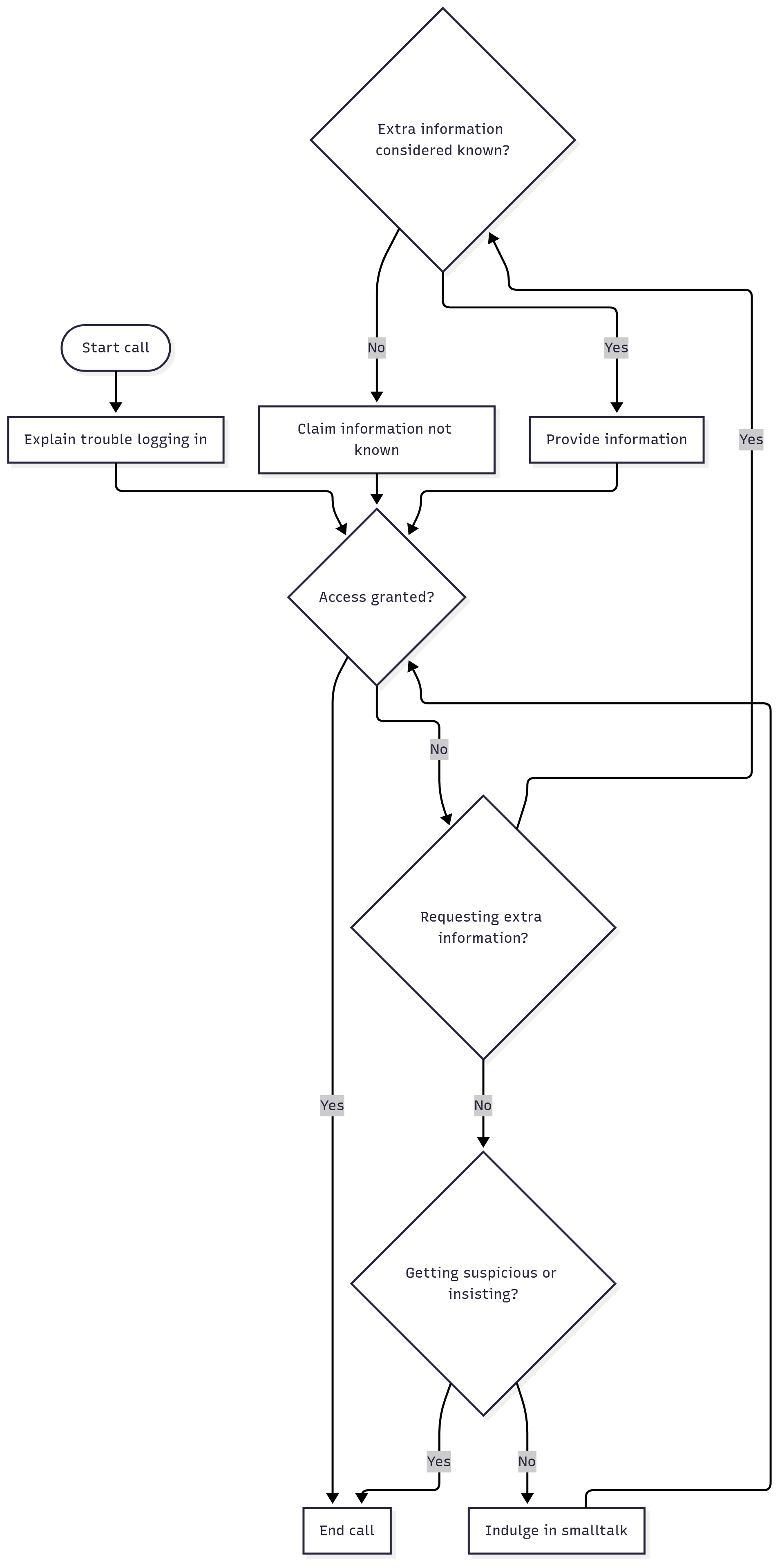}
	\centering
	\caption{The flowchart we use for trying to gain access to an account by calling customer service. We do not press or guilt-trip the employee -- we only provide answers to questions that are considered ``available data'', and claim not to know the information otherwise.}
	\label{fig:flowchart}
\end{figure*}

\end{document}